\documentclass[prb,aps,showpacs,a4paper,floatfix]{revtex4}
\usepackage{amsmath,textcomp,graphicx}
\usepackage{times}              
\usepackage[OT1,PL]{ogonek}     
\begin{document}
\title{%
Spontaneous ordering as an intrinsic effect at the mesoscopic scale: 
A vibrational insight \linebreak
in (Zn,Be)Se by Raman scattering and first-principles calculations 
}
\author{O. Pag\`{e}s,\footnote{%
Author to whom correspondence should be addressed.
Electronic Address: pages@univ-metz.fr}
A.~V. Postnikov, A. Chafi}
\affiliation{LPMD, Universit\'{e}
de Metz, 1 Bd. Arago, F-57078 Metz, France}
\author{D. Bormann, P. Simon}
\affiliation{CRMHT, CNRS UPR4212, Universit\'{e} d'Orl\'{e}ans,
1D Av. de la Recherche Scientifique, 45071 Orl\'{e}ans, France}
\author{F. Firszt}
\affiliation{Institute of Physics, N. Copernicus University,
Grudzi\k{a}dzka 5/7, 87-100 Toru\'{n}, Poland}
\author{W. Paszkowicz}
\affiliation{Institute of Physics, Polish Academy of Sciences,
Al. Lotnik\'{o}w 32/46, 02-668 Warsaw, Poland}
\author{E. Tourni\'e}
\affiliation{CEM2, CNRS UMR5507, Universit\'e Montpellier 2,
34095 Montpellier, France}
\begin{abstract}
The recent finding of a 1-bond$\rightarrow$2-phonon `percolation'-type 
behaviour in several random zincblende alloys, supporting an unsuspected 
1-bond$\rightarrow$2-mode behaviour in the bond length distribution, renews 
interest for a discussion of CuPt-type spontaneous ordering (CPSO) 
as a purely intrinsic effect. We investigate this key issue from 
both experimental (Raman scattering) and theoretical (first-principles 
bond length and phonon calculations) sides, focusing on Zn$_{1-x}$Be$_x$Se 
($0<x=0.92$). Its large contrast in cation-anion bond 
lengths should stimulate CPSO to some extent, and is at the origin 
of a uniquely well-resolved 1-bond$\rightarrow$2-phonon behaviour for the 
Be-Se species. Based on percolation, CPSO finds a previously 
missing natural explanation, at the mesosocopic scale. Also, 
the transfer of oscillator strength from the high- to the low-frequency 
Be-Se mode in the Raman spectra potentially emerges as a sensitive 
probe of CPSO in ZnBeSe. First-principles calculations indicate 
that the transfer is completed for the value of the long range 
order parameter of $\eta\sim$0.5 in ZnBeSe$_2$ ($x\sim$0.5), 
corresponding to a pure 1-bond$\rightarrow$1-mode behaviour. With further 
ordering the alloy is forced to re-adopt a non-rewarding 
1-bond$\rightarrow$2-mode behaviour, due to the formation of ZnSe/BeSe 
micro-domains in the crystal. Therefore $\eta\sim$0.5 appears as an intrinsic 
limit to CPSO in ZnBeSe$_2$, and possibly in stoichiometric alloys in general. 
\end{abstract} 
\maketitle

\section{INTRODUCTION}
The current believe so far is that in random A$_{1-x}$B$_x$C 
zincblende semiconductor alloys, where C denotes indifferently 
the cationic or the anionic species, the A-C and B-C bonds tend 
at any $x$ value to keep their natural lengths (L) as inherited 
from the parent compounds.$^1$ 
This we refer to as the 1-bond$\rightarrow$1-mode (L) behaviour 
in the bond length distribution. Only a slight 
adaptation occurs when $x$ varies, due to the necessity to fit 
into the lattice, which, on the average, shrinks or dilates. 
With this, the contrast between the A-C and B-C bond lengths 
remains always smaller in the alloy than between the parent compounds 
(see, e.g., Fig.~2 in Ref.~1). In its impact on the lattice 
dynamics, the 1-bond$\rightarrow$1-mode (L) situation manifests itself as 
a typical 1-bond$\rightarrow$1-mode behaviour in the transverse optical 
(TO) Raman spectra, as accounted for by the 
well-known modified-random-element-isodisplacement 
(MREI) model of Chang and Mitra.$^2$ This is based on the virtual 
crystal approximation that averages the alloy disorder locally, 
corresponding \textit{in fine} to a description of the alloy 
at the \textit{macroscopic} 
scale, i.e. in terms of an effective medium. Note that, usually, 
the larger the bond length (L), the smaller the bond force constant, 
and hence the lower the vibration frequency of the phonon mode (TO).

The 1-bond$\rightarrow$1-mode (L) picture is especially problematic with 
respect to the discussion of spontaneous ordering (SO) in alloys. 
First, it implies that at a given $x$ value the bond length 
does not depend on the local neighbourhood of the individual 
bonds. Accordingly no benefit, i.e. no minimization of the local 
strain energy due to the bond length mismatch, can be expected 
if a substitution of A by B be not purely random. Second, the 
mechanism of SO as usually observed in stoichiometric alloys ($x$=0.5), 
referred to as ABC$_2$ for simplicity hereafter, is counter-intuitive when 
starting from the 1-bond$\rightarrow$1-mode (L) description of the 
lattice relaxation in a random alloy. We recall that, most frequently, 
SO in ABC$_2$ consists of the formation of a 
A$_{0.5(1-\eta)}$B$_{0.5(1+\eta)}$C/A$_{0.5(1+\eta)}$B$_{0.5(1-\eta)}$C 
quasi-superlattice along the $[{\bar 1}11]$ and $[1 {\bar 1} 1]$
crystal directions,$^3$ where $\eta$, the long-range order parameter, 
measures the average deviation from equal representation of A and B species 
in the (111) substituting planes, as expected in a random alloy.$^{3,4}$ 
This is currently referred to as CuPt-b type ordering. Here the consecutive 
(A,B) substituting planes have, in alternation, an excess and deficiency 
of A (or B). The two limiting $\eta$ values correspond to the random 
substitution ($\eta$=0) and to the perfect CuPt-b ordering ($\eta$=1). 
In other words, the SO generates a local segregation of 
the A and B species within a sequence of intercalated planes, 
i.e. it enforces the building up of regions which, locally, resemble 
the parent compounds. In the spirit of the 1-bond$\rightarrow$1-mode (L) 
description this should enlarge the overall contrast between 
the A-C and B-C bond lengths in the alloy, as discussed above, 
and thereby reduce the crystal stability. 

In summary when starting from the traditional 1-bond$\rightarrow$1-mode 
(L) description of the lattice relaxation in a random mixed crystal 
at the macroscopic scale, not only would we fail to understand 
intuitively ($i$) how SO can be rewarding for the crystal at all, 
but also ($ii$) why SO takes specifically the form of local segregation 
of the A and B species into alternate substituting planes, with 
interleaving planes of C atoms. 

Certainly the whole of this has discouraged a discussion of the 
CuPt-b type SO (CPSO) in semiconductor alloys in terms of an \textit{intrinsic} 
effect, i.e. the result of an optimal lattice relaxation in the \textit{bulk} 
of the crystal with respect to minimization of the local strain 
energy as due to the contrast between the A-C and B-C bond lengths, 
and pushed towards a discussion more in terms of an \textit{extrinsic} 
effect, i.e. in relation to the growth conditions at the \textit{surface}.$^5$

Actually the growth conditions play an important role. For a 
brief survey we focus on InGaP$_2$ that has been extensively studied 
in the literature, both experimentally and theoretically (for 
a detailed review see Ref.~6). First it has been shown that the growth rate, 
the growth temperature, and also the III-V ratio do influence the 
\textit{degree} of CuPt-b ordering, i.e. the $\eta$ value (see Ref.~3, 
and Refs. therein). Further, it was shown that substrate misorientation over 
a few degrees towards [111] may enforce massive ordering along one only 
of the two possible $[\bar{1}11]$ and $[1\bar{1}1]$ directions, leading to 
a so-called single-variant CuPt-b type ordering.$^{3,4}$ 
Besides, first-principles calculations performed by Zhang et al.$^7$ indicate 
that, depending on the orientation of the surface phosphorus dimers during 
the growth process, the \textit{nature} of ordering may change from 
the standard CuPt-b form to the much less frequent CuPt-a one, corresponding 
to the building up of a monolayer superlattice along the alternative 
$[111]$ and $[\bar{1}\bar{1}1]$ crystal directions, or even result in 
a triple-period ordering. Experimentally this was evidenced by 
Gomyo \textit{et al.}$^8$ The underlying mechanism is surfacial: the ordering 
basically results from (Ga,In)-site selectivity minimizing the subsurface 
strain induced by the top surface P-dimers during the growth process.$^7$
This has lead to an idea to influence the ordering in depth 
of III-V alloys by using specific impurity atoms selected so 
as to stay on the top surface of the alloy during its deposition. 
In particular, at present much attention is awarded to the fact 
that such so-called surfactant atoms may stimulate or even block the natural 
trend of the alloy to develop the CPSO (see Ref.~9 and references therein). 

One remarkable point about the CPSO is an apparent limit the 
nature has set to its magnitude. The highest observed value of 
$\eta$ is $\sim$0.5 (in InGaP$_2$ see Fig.~2 in Ref.~10). We are 
not aware of any ABC$_2$ alloy, where a higher degree of SO could 
ever be achieved.

Of course the degree of ordering can be freely chosen in a calculation. 
Zhang et al.$^{10}$, in their sequence of (empirical potentials-based) 
lattice relaxations in large supercells, taking average over 
many different configurations for each $\eta$ value, deduced 
that the character of bond length distributions in both Ga-P 
and In-P species basically changes near $\eta\sim$0.5. The authors 
note the closeness of this `anomaly' 
to the experimentally achievable limit of ordering. However, 
they did not elaborate what exactly in their bond lengths distributions 
beyond $\eta\sim$0.5 would make such configurations difficult to occur 
in reality. From this, we deduce that the limit of SO at
$\eta\sim$0.5 has to do with the lattice relaxation in the bulk 
of the crystal, as due to the contrast between the A-C and B-C 
bond lengths, and not only with the lattice relaxation at the 
surface of the alloy during its deposition, as conditioned by 
the growth parameters and/or surfactants.$^9$ What emerges is 
that $\eta\sim$0.5 should be discussed in terms of an \textit{intrinsic} 
limit to CPSO in GaInP$_2$. This appeals to rethink the whole 
scenario of CPSO in terms of a purely \textit{intrinsic} effect, which 
has attracted little attention so far. 

We emphasize that for doing so, one should look into more fine 
details of lattice relaxation in random alloys than those covered 
by the traditional, and crude, 1-bond$\rightarrow$1-mode (L) picture at 
the macroscopic scale, that fails to account for the CPSO as 
already discussed.

In the last few years we have shown that several A$_{1-x}$B$_x$C 
random alloys, e.g. (Zn,Be) chalcogenides and (Ga,In)-\textit{V} systems 
(where \textit{V} stands for As or P), exhibit in fact 
a 1-bond$\rightarrow$2-mode behaviour in the TO Raman spectra.$^{11}$ 
The microscopic mechanism for the 1-bond$\rightarrow$2-mode (TO) Raman 
behaviour was pinpointed 
by first-principles/atomistic calculations to the difference 
in bond length (L) according to whether the bonds stay within 
the Zn/In-rich region or within the Be/Ga-rich one, that result 
from natural fluctuations in the alloy composition at the local 
scale$^{12}$ This introduces a description of the lattice dynamics/relaxation 
in a semiconductor mixed crystal A$_{1-x}$B$_x$C at the \textit{mesoscopic} 
scale, in terms of a composite made of the two finely interlaced 
A-rich and B-rich regions. 

One key point is that, \textit{for a given bond species, the bonds 
appear to be longer (shorter) within the host region that refers 
to the parent material with the smaller (larger) lattice constant}. 
For example, the Be-chalcogen bonds are longer in Be-rich than 
in Zn-rich regions of (Zn,Be) chalcogenides, and the Ga-based 
bonds are shorter in In-rich than in Ga-rich regions of (Ga,In)-based 
systems. More generally, if B-C refers to the shortest bond length 
in A$_{1-x}$B$_x$C alloys, we distinguish for each of the A-C and 
B-C bond species between the `short' bonds from the A-rich region, 
that vibrate at a higher frequency, and the `long' bonds from 
the B-rich region, that vibrate at a lower frequency. The phonon 
modes (TO) and bond lengths (L) are labelled as 
(TO,L)$_{\rm (A,B)-C}^{\rm A,B}$, where the subscript refers to the bond 
species, and superscript indicates the A-rich or B-rich host region. 
A typical scattering 
of lengths, within a given bond species (A-C) or (B-C), depending 
on their affiliation to A-rich or B-rich host region, is of the 
order of the percent ($\delta\sim$1\%), but amongst the 
short B-C bond species this scattering is larger, thereby corresponding 
to a clear separation between the phonon lines $\Delta$. For 
the long A-C bond species, \ensuremath{\Delta} effectively vanishes. The 
full description of the 1-bond$\rightarrow$2-mode (TO) behaviour, including 
the dependencies of the frequency and the strength of each individual 
mode on the alloy composition, was finalized into a so-called 
`percolation' model. Details are reported elsewhere.$^{11}$ We reproduce 
in Fig.~1 the resulting scheme of phonon modes for the (Zn,Be)-chalcogenides 
in the random case, of present interest here, to help us in the 
discussion. The 1-bond$\rightarrow$1-mode (TO) MREI-like behaviour 
is also indicated there by dotted lines. The real curve for 
Zn$_{1-x}$Be$_x$Se is displayed e.g. in Fig.~1 of Ref.~11. 

\begin{figure}[b]
\centerline{\includegraphics[height=0.5\textwidth,angle=0]{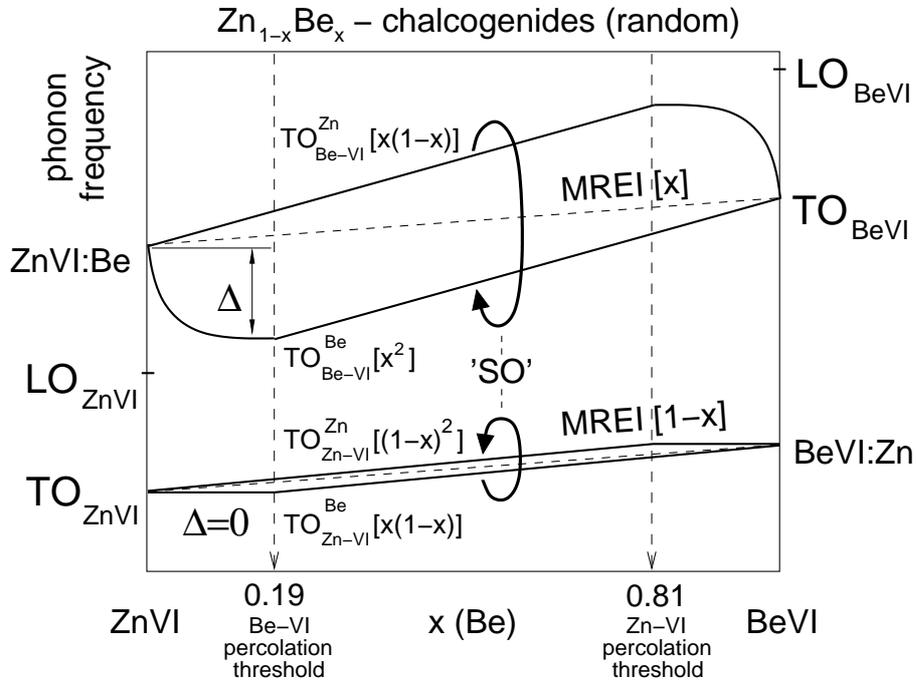}}
\caption{%
Schematic view of the 1-bond$\rightarrow$2-mode (TO) `percolation' model
for Zn$_{1-x}$Be$_x$ chalcogenides (solid lines) in the random case. The dotted
arrows indicate the Be-Se and Zn-Se bond percolation thresholds, corresponding
to critical phonon behaviors. The traditional 1-bond$\rightarrow$1-mode (TO)
MREI-like model is also represented for reference purpose (diagonal dashed
lines). The impurity mode of Be (Zn) in ZnVI (BeVI) at $x{\sim}0$ ($x{\sim}1$)
is referred to as ZnVI:Be (BeVI:Zn). The strength of the individual modes
depending on the alloy composition is indicated within square brackets
for each model. The curved arrows indicate the antagonist effects of SO
within the Zn-VI and Be-VI phonon double branches.}
\end{figure}

Clearly the 1-bond$\rightarrow$2-mode (TO,L) description of the lattice 
dynamics/relaxation in random alloys provides a more attractive area 
than the standard 1-bond$\rightarrow$1-mode (TO,L) alternative for 
the intuitive discussion of the above key issues ($i$)--($ii$) about the SO. 

The preliminary results that we have obtained so far with GaInP$_2$ are 
the following. First we could explain the puzzling evolution 
of the Raman/infrared spectra with increasing order in terms 
of reinforcement of the low-frequency TO$_{\rm Ga-P}^{\rm Ga}$
mode at the cost of the high-frequency TO$_{\rm Ga-P}^{\rm In}$ mode. 
Therefore, at least in principle, \textit{the strength ratio 
$R$ between the low- and high-frequency Ga-P modes in their 
1-bond$\rightarrow$2-mode 
description can be used as a probe of the CPSO.} Basically the 
larger the R value, the higher the degree of CPSO. Transposed 
to bond lengths this indicates that the SO would tend to generate 
those local atom arrangements in the crystal that favour the 
long Ga-P bonds from the Ga-rich region to the detriment of the 
short Ga-P bonds from the In-rich region. This way the contrast 
in bond lengths between the short Ga-P bonds and the long In-P 
bonds is minimized, resulting in higher crystal stability. Note 
that we discuss SO in GaInP$_2$ as a purely intrinsic effect, 
which relies entirely on the contrast between the Ga-P and In-P 
bond lengths. Further, we could provide a preliminary explanation 
as to why CPSO is apparently limited to $\eta\sim$0.5 in GaInP$_2$.$^{13}$ 
Basically the reason is that the 
short Ga-P bonds from the In-rich region have totally disappeared 
at this limit, as can be inferred from total disappearance of 
the TO$_{\rm Ga-P}^{\rm In}$ mode in the Raman/Infrared spectra,$^{14,15}$ 
which suppresses the driving force for further ordering.

However, many key questions remain open. The discussion above 
offers some insight into the issue ($i$), that is, how SO can be 
rewarding for a crystal. Yet the issue ($ii$) remains unaddressed, 
because no link was established between the Raman spectra and 
underlying details of the crystal structure depending on $\eta$. 
In particular, the reason as why the CuPt type atom arrangement 
would lead to disappearance of the shortest Ga-P bonds was not 
clarified so far. Besides, a novel question emerges: ($iii$) if 
CPSO is actually hindered beyond $\eta\sim$0.5, as we proposed for GaInP$_2$, 
what is the forbidding mechanism behind at the microscopic scale? 

In the present work we address the above important issues with 
the aim to derive a consistent picture for the CPSO that incorporates 
the prerequisite of a 1-bond$\rightarrow$2-mode (TO,L) description 
of the lattice relaxation at $\eta$=0 for both the Ga-P and In-P species. 
Ideally the key notion of an intrinsic limit to CPSO, as earlier discussed 
for the reference GaInP$_2$ system, should come out. This we investigate 
via a thorough study of the dependence of the 1-bond$\rightarrow$2-mode 
(TO,L) behaviour versus ordering in the Zn$_{1-x}$Be$_x$Se system, from 
both the experimental and theoretical sides. 

An immediate motivation for focusing on this system is that our 
earlier Raman data seem to indicate CPSO to some extent in our 
set of Zn$_{1-x}$Be$_x$Se epilayers. For example, at the stoichiometry 
($x$=0.5) the strength ratio $R$ between the TO$_{\rm Be-VI}^{\rm Be}$
(at low frequency) and the TO$_{\rm Be-VI}^{\rm Zn}$
 (at high frequency) modes should be close to 1 in case of random 
Be substitution for Zn in (Zn,Be) chalcogenides (refer to Fig.~ 1). 
While this is true for ZnBeTe$_2$ (see Fig.~3 of Ref. 16), 
$R$ is larger than expected in ZnBeSe$_2$ (see Fig.~4 of Ref. 7). 
In fact, in Zn$_{1-x}$Be$_x$Se epilayers $R{\sim}$1 at $x{\sim}$0.4 
(see Fig.~2 of Ref. 11) and not at $x{\sim}$0.5. More generally, 
the $R$ value is always smaller in Zn$_{1-x}$Be$_x$Se than 
in Zn$_{1-x}$Be$_x$Te, at any Be content $x$. For an insight at large 
$x$ values, where the Be-based signal shows up clearly, compare the top curves 
from Fig.~2(c) in Ref. 16 and from Fig.~4(c) in Ref. 11. 
The apparent $R$-anomaly in Zn$_{1-x}$Be$_x$Se has first been tentatively 
attributed to larger damping or smaller Faust-Henry coefficient 
of the Be-Se vibrations in the Zn-rich region, for an unspecified 
intrinsic reason (see Ref.~11, and Refs. therein). More consistently, 
another possible explanation is that our Zn$_{1-x}$Be$_x$Se epilayers 
exhibit CPSO to some extent, by an analogy with InGaP$_2$ (refer 
above). In this case the apparent $R$-anomaly would not be intrinsic. 
In particular, it should depend on the growth conditions to some 
extent. We investigate this issue in the first, experimental, 
part of this work by using an alternative set of Zn$_{1-x}$Be$_x$Se 
alloys, grown as single crystals, with the underlying idea that 
these might be less prone to SO than the epilayers. Indeed while 
re-examining the Raman/infrared data available for GaInP$_2$ in 
the literature, we noted that only those samples grown as epitaxial 
layers do significantly exhibit the CPSO. The samples grown as 
single crystals seem to be random (refer to the inset of Fig.~2 in Ref. 13). 
Apparently this is not fortuitous, as the same trend was also evidenced 
with ZnSe$_x$Te$_{1-x}$, over the whole composition range.$^{17}$

We mention that the amount of SO in our specific Zn$_{1-x}$Be$_x$Se 
epilayers, if any, is expected to be low. Indeed we have detected 
earlier a singularity in the $x$-dependence of the frequency of 
the TO$_{\rm Be-Se}^{\rm Be}$ mode, that refers to the Be-rich region, 
when the latter region turns from a dispersion of bounded clusters 
with fixed internal geometry (small $x$-values, stable frequency) 
into a pseudocontinuum with smoothly $x$-dependent internal geometry 
(large $x$-values, smoothly $x$-dependent frequency). The transition occurs 
at the so-called Be-Se bond percolation threshold, denoted $x_{\rm Be-VI}$. 
Remarkably the singularity was observed at $x_{\rm Be-VI}\sim$0.19 
(refer to Fig.~1 in Ref. 11, for example),  i.e. the theoretical 
$x_{\rm Be-VI}$ value if random B substitution to A is assumed on the 
fcc (A,B)-sublattice of zincblende A$_{1-x}$B$_x$C alloys [18]. On 
this basis our Zn$_{1-x}$Be$_x$Se epilayers were previously attested 
as quasi-random systems, hence their $\eta$ values should remain close to zero.

More generally ZnBeSe is a choice system to address the key questions 
($i$) to ($iii$) above. First it exhibits a contrast in the bond 
lengths ($\sim$9\%) comparable to that in GaInP ($\sim$7\%). 
Accordingly if CuPt ordering has an intrinsic origin and stems 
from the latter contrast, as we expect, this should stimulate 
CPSO to some extent, depending on the growth conditions. Second, 
the contrast in the bond force constants is extremely large in 
ZnBeSe, essentially due to the unique covalent character of the 
Be-bonding among II-VI compounds.$^{19,20}$ A consequence is that 
the Zn- and Be-based bonds vibrate in well-separated frequency 
domains, which allows, in particular, a direct insight into the 
Be-based spectral range, where the 1-bond$\rightarrow$2-mode (TO) behaviour 
is especially pronounced. An uniquely good resolution of split 
Be-Se lines ($\Delta{\sim}$50~cm$^{-1}$) allows reliable insight 
into the relative strength of the two Be-Se modes, that seems 
to be a relevant probe of the CPSO by analogy with GaInP$_2$.$^{13}$ 
In fact ZnBeSe seems to be a more favourable test case than InGaP$_2$, 
in which the Ga-P and In-P spectral ranges do overlap strongly, 
and the splitting between the two Ga-P TO modes is two times 
smaller ($\Delta{\sim}$20~cm$^{-1}$). 

We are not aware of any vibrational study of the CPSO within 
the classes of II-VI or I-VII alloys. As for III-V's, little 
work has been dedicated to the dependence of lattice dynamics 
versus ordering in systems other than GaInP$_2$, in spite of the 
fact that the tendency for CPSO is a common feature in this class 
of alloys. Based on Raman spectroscopy data, an evidence for 
CPSO has been reported in GaInAs$_2$ grown lattice-matched to 
(001) InP substrates,$^{21,22}$ and also in GaAs$_{1-x}$N$_x$ 
($x{\sim}$ 0-5\%).$^{23}$ We note that in all cases the vibrational insight 
into the CPSO remained but qualitative. 

More precisely we note that the previous attempts to clarify 
the influence of ordering on the Raman spectra of the above systems 
were mostly performed in the longitudinal optical (LO) symmetry. 
Here, the most reliable indicator of SO seems to be the emergence 
of the folded longitudinal acoustical (FLA) phonon, which otherwise 
does not show up in the Raman spectrum of the random alloy.$^{22}$ 
Unfortunately there does not seem to exist any straightforward 
correlation between the strength of this mode and the degree 
of ordering. By way of example, in the case of GaInP$_2$ an unambiguous 
emergence of the FLA mode at $\sim$205~cm$^{-1}$ requires a 
minimum amount of ordering of $\eta{\sim}$0.3, and the strength 
of the mode remains approximately constant at larger $\eta$
values up to $\sim$0.5 (refer to Fig.~5 in p.~399 of Ref.~6, 
and to Fig.~4 in Ref.~3). As far as pure optical modes 
are concerned, we have shown elsewhere in extensive detail that 
the TO symmetry is much more sensitive than the LO alternative 
to detect possible changes in either the frequency or the oscillator 
strength of the individual oscillators in the crystal.$^{11}$ With 
respect to earlier reported measurements in the TO symmetry, 
we are only aware of important contributions by Mestres 
\emph{et al.}$^{24}$ and Cheong \emph{et al.}$^{14}$ dedicated to GaInP$_2$,
but the discussion remained qualitative only. Taken together, this 
reinforces the motivation for the present detailed investigation 
of the lattice dynamics of Zn$_{1-x}$Be$_x$Se in the TO symmetry.

The manuscript is organized as follows. In Sec. II we give the 
experimental details concerning the preparation of the samples 
and the recording of the Raman spectra, and we outline the first-principles 
methods for the calculations of the bond length and phonon properties. 
As an introduction to Sec. III we propose a consistent intrinsic 
mechanism for the CPSO, based on our 1-bond$\rightarrow$2-mode (TO,L) 
description of the lattice dynamics/relaxation in random mixed crystals at 
the mesoscopic scale. According to this, CPSO should show up 
via a systematic transfer of oscillator strength between the 
two modes that refer to the same bond species in the Raman spectra. 
By comparing the well-resolved 1-bond$\rightarrow$2-mode (TO) 
Be-Se Raman responses from Zn$_{1-x}$Be$_x$Se single crystals 
(0.10${\leq}x{\leq}$0.53) and epitaxial layers (0${<}x{\leq}$0.92), 
such a transfer is actually evidenced for the epilayer set. This corresponds 
to a slight over-representation of the `long' Be-Se bonds from the Be-rich 
region, to the detriment of the `short' Be-Se bonds from the 
Zn-rich region, as should be expected in the case of moderate 
CPSO. Quantitative information is derived from contour modeling 
of the TO Raman lineshapes by using our phenomenological `percolation' 
model. In Sec. IV we perform first-principles calculations with 
a series of fully-relaxed 32-atom Zn$_8$Be$_8$Se$_{16}$ supercells 
for complementary insight at the microscopic scale. First this 
is to verify that CuPt type ordering actually generates the interplay 
between the two Be-Se Raman modes as observed with the epilayer 
set. The calculations are pushed further to provide insight into 
the critical $\eta$ value corresponding to the full disappearance 
of the `short' (`long') Be-Se (Zn-Se) species in the representative 
ZnBeSe$_2$ alloy. At last, we investigate the nature of the threshold 
mechanism, if any, that should hinder SO beyond this limit, as earlier 
proposed for GaInP$_2$.$^{13}$ Concluding remarks are given in Sec. V. 

\section{EXPERIMENT AND FIRST-PRINCIPLES METHOD}
We analyzed two sets of Zn$_{1-x}$Be$_x$Se samples, i.e. 
$\sim$1~$\mu$m-thick layers (0${<}x{\leq}$0.92) grown by molecular beam 
epitaxy onto a (001) GaAs substrate using several ZnSe monolayers as 
an intermediate buffer, and single crystals ($x{\sim}$0.10, $\sim$0.25, 
$\sim$0.50) grown as cylinders (8 mm in diameter, 20 mm in length) by using 
the high-pressure Bridgman method. The single crystals were oriented 
and cut perpendicular to their growth direction, that was checked 
to be the [111] crystal axis by high-resolution X-ray diffraction, 
and then mechanically polished to optical quality. For each sample 
the $x$ value was determined with a typical accuracy of $\pm$0.5\% 
from the lattice constant measured by X-ray diffraction, assuming 
a linear dependence (Vegard's law). The epilayers are lattice-matched 
with the GaAs substrate at $x{\sim}$0.03. Combined X-ray measurements 
of the in-plane and out-of-plane lattice constants revealed a 
substrate-induced residual biaxial tensile strain in the layers 
up to $x{\sim}$0.25, above which limit the layers are fully 
relaxed.$^{25}$ We have checked earlier that the residual strain 
in the epilayers is homogeneous, at least from the vibrational 
side (refer to Fig.~4 in Ref.~26). Basically there is no such 
variation of the residual strain in the epilayer from the ZnBeSe/GaAs 
interface to the top surface that could cause any detectable 
distortion of either Zn-Se or Be-Se phonon lineshapes, apart 
from slight overall red shifts. Besides, micro-Raman spectra 
recorded along the $\sim$1{\textperthousand} (100) slope of a bevelled 
edge realized by chemical etching with the Zn$_{0.38}$Be$_{0.62}$Se/GaAs 
system have shown that the intermediary ZnSe buffer layer has 
initiated the growth of a thin ZnSe-like amorphous layer prior 
the proper deposition of the alloy at the nominal composition.$^{27}$ 
We deduce that the growth of the alloy is free from the 
substrate influence at large $x$ values. We retain that there are 
two well-separated $x$ values corresponding to critical behaviours 
with respect to the substrate-induced tensile strain in the epilayers, 
i.e., $x{\sim}$0.25 and $\sim$0.6. 

Raman spectra were recorded at room temperature with the Dilor 
microprobe setup by using the 514.5 nm Ar$^{+}$ excitation. Pure 
TO spectra from the epilayers were obtained in backscattering 
on the (110) edge face with unpolarized excitation/detection 
(geometry 1, see Fig.~2)$^{28}$ By using the ($\times$100) microscope 
objective the laser spot on the sample surface could be reduced 
down to $\sim$1~${\mu}$m, but the microprobe generally overlaps 
onto the GaAs side, which activates the parasitic 
$TO_{\rm GaAs}$ ($\sim$268~cm$^{-1}$, allowed) and 
$LO_{\rm GaAs}$ ($\sim$292~cm$^{-1}$, theoretically forbidden) modes. 
For the single crystals we used (111) backscattering, corresponding to 
both TO and LO modes allowed. Specific TO access was achieved 
by taking crossed polarizations for the incident and scattered 
beams, and turning the sample around the [111] axis until the 
polarizations got parallel to the $\left[ 1{\bar 1}0 \right]$
and $\left[1 1 {\bar 2} \right]$ crystal axis, respectively 
(geometry 2, see Fig.~2).$^{28}$ 

The first-principles calculations of bond lengths and phonons 
were performed for the stoichiometric composition using a series 
of Zn$_8$Be$_8$Se$_{16}$ supercells constructed so as to represent 
different values of the order parameter, $\eta$=0 through 1, by using 
the computer code {\sc Siesta}.$^{29,30}$ The details concerning the building 
up of the supercells are reported in Sec. IV. These are fully relaxed, 
with respect to crystal cell parameters and internal atomic positions. 
The norm-conserving pseudopotentials and the basis set are just those 
already optimized for our earlier calculations with the ordered 
Zn$_{4-n}$Be$_n$Se$_4$, $n$=1,3 ($x$=0.25, 0.75) supercells and our prototype 
disordered Zn$_{26}$Be$_6$Se$_{32}$ supercell at the Be-Se bond percolation 
threshold. Extensive information is given in Ref.~12. 

\section{RAMAN SCATTERING AND `PERCOLATION' MODEL: 
AN INSIGHT AT THE MESOSCOPIC SCALE}
As a basis for the discussion of the experimental data in Sec.~III-B, 
we outline in Sec.~III-A a more accomplished version of 
the mechanism earlier introduced to explain the puzzling evolution 
of the 1-bond$\rightarrow$2-mode (TO) Ga-P behaviour in the course of 
varying the order parameter $\eta$ in the reference GaInP$_2$ system.$^{13}$
In particular this version fills the gap regarding question ($iii$) raised 
in Sec.~I. \textit{A priori} the same picture should apply to the similar 
1-bond$\rightarrow$2-mode (TO,L) Be-Se behaviour in Zn$_{1-x}$Be$_x$Se, 
by analogy. 

\subsection{CPSO: An intrinsic mechanism at the mesoscopic scale}
Clearly the existence of two different bond lengths (L) per species 
in a random A$_{1-x}$B$_x$C alloys ($\eta$=0), supported by the 
1-bond$\rightarrow$2-mode (TO) `percolation' type behaviour of the lattice 
dynamics, is not optimum regarding the stability of the crystal structure. 
A minimization of the local strain energy, hence higher crystal stability, 
can be achieved simply by reducing the diversity in bond length in the sense 
of just selecting the one single bond length per species that 
minimizes the contrast between the A-C (long) and B-C (short) 
bond lengths, i.e. the `shorter' A-C bonds from the A-rich region 
and the `longer' B-C bonds from the B-rich region. This explains 
how SO can actually be rewarding for the crystal -- recall issue 
($i$) in Sec. I. Ultimately only the latter bond species should 
remain in the crystal. Transposed to phonon spectra, SO would 
correspond to reinforcement of the TO$_{\rm A-C}^{\rm A}$ and 
TO$_{\rm B-C}^{\rm B}$ modes at the cost of the TO$_{\rm A-C}^{\rm B}$
and TO$_{\rm B-C}^{\rm A}$ modes, respectively, to full disappearance 
of the latter (refer to the antagonist curved arrows in Fig.~1 in the case 
of ZnBeSe). 

An interesting question -- in reference to issue ($ii$) of Sec.~I -- is 
which local atom arrangements would best promote this tendency? Apparently 
those leading to local phase separation, because this is the 
\textit{sine qua non} condition to confine B-C and A-C bonds within B-rich 
and A-rich environments, respectively. When starting from a situation where 
the A and B atom species are represented in each substituting plane 
in proportion to the alloy composition ($\eta$=0), certainly the most 
straightforward way to achieve this would be massive Zn$\leftrightarrow$Be 
exchange between adjacent (111) substituting planes so as to enforce 
their alternate A- and B-enrichment throughout the crystal. In the 
representative ABC$_2$ system this eventually results in long range order 
under the form of a 
A$_{0.5(1+\eta)}$B$_{0.5(1-\eta)}$C/A$_{0.5(1-\eta)}$B$_{0.5(1+\eta)}$C 
superlattice along the [111] crystal direction.

As the mechanism that we propose for the CPSO in A$_{1-x}$B$_x$C 
mixed crystals entirely relies upon the contrast in bond lengths 
between the A-C and B-C species, the CPSO should not be observed 
when the latter contrast vanishes. As a matter of fact, it is 
well-known that Al$_{1-x}$Ga$_x$As, which is the ultimate system 
with this respect (the difference between the Al-As and Ga-As 
bond lengths is smaller than $\sim$1.5{\textperthousand}), never develops 
the CPSO.$^{6}$

\begin{figure}[t]
\centerline{%
\includegraphics[height=0.7\textwidth,clip=true,angle=270]{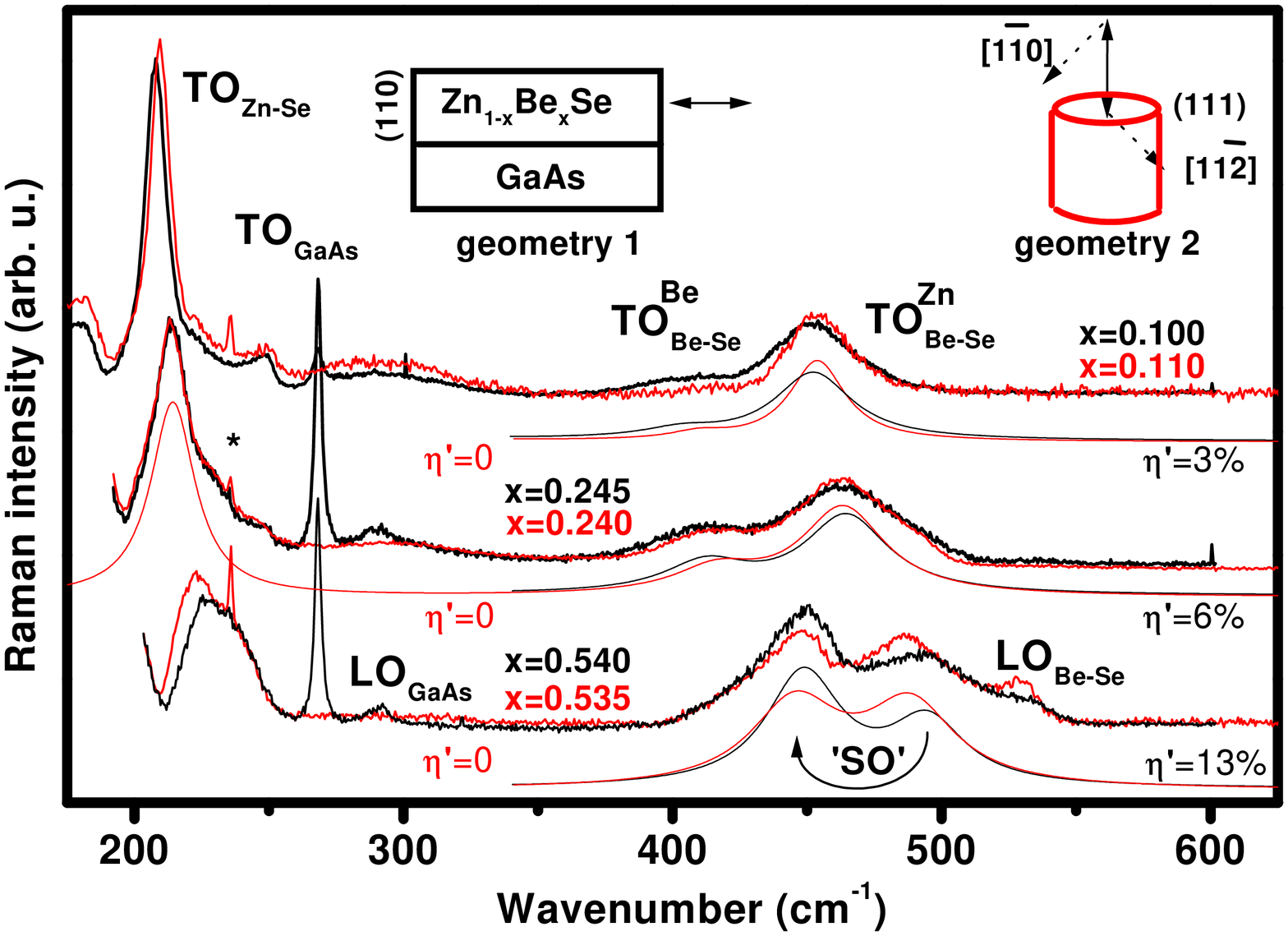}
}
\caption{%
(color online): Superimposed TO Raman spectra from the Zn$_{1-x}$Be$_x$Se
single crystals (red) and epilayers (black) with similar $x$ values within
$\sim$1\%, after normalization to the strength of the TO$_{\rm Zn-Se}$
signal, taken as an internal reference. The corresponding $x$ values and
scattering geometries are distinguished by using the same color code.
The double (dotted) arrows indicate the excitation-detection
(polarization) direction(s). The multi-mode TO curves obtained
from contour modeling of the Raman spectra by using a three-oscillator
version [Zn-Se, (Be-Se)$^{\rm Be}$, (Be-Se)$^{\rm Zn}$] of the `percolation'
model are shown below the data, while using the same color code,
with special emphasis upon the sensitive Be-Se spectral range.
An example of theoretical curve covering both the Zn-Se and Be-Se
spectral ranges is shown at intermediary $x$ value. In each case
the corresponding $\eta'$ values are indicated, where
$\eta'$ measures the transfer of Be-Se oscillator strength as indicated
by the curved arrow, the result of SO (refer also to the upper curved arrow
in Fig. 1). The asterisk marks a parasitical laser line.}
\end{figure}

Thus, when starting from the 1-bond$ßrightarrow$2-mode (TO,L) description 
of random A$_{1-x}$B$_x$C alloys ($\eta$=0), an intrinsic driving force 
behind SO is identified, in response to question ($i$), and it becomes clear 
why the mechanism of SO consists of local segregation of the A and B species 
within alternate substituting planes, that is the answer to question ($iii$). 
Below we discuss on this novel basis the apparent $R$-anomaly earlier 
detected in the Be-Se TO Raman response of the Zn$_{1-x}$Be$_x$Se 
epilayers (refer to Sec. I).

\subsection{Experimental results and discussion 
(Zn$_{1-x}$Be$_x$Se, $0{\leq}{x}{\leq}0.92$)}
We superimpose in Fig.~2 the raw TO Raman spectra from the single 
crystals (red) and the epilayers (black) with $x$ values similar 
within 1\%, after normalization to the strength 
of the 1-bond$\rightarrow$1-mode (TO) MREI-like 
TO$_{\rm Zn-Se}$ signal at $\sim$215~cm$^{-1}$, taken as an internal reference. 
The Be-Se bonds vibrate at higher frequency, near 450~cm$^{-1}$, due to 
their stiffer/shorter character and their smaller reduced mass. Both the 
TO$_{\rm Be-Se}^{\rm Be}$ mode, at $\sim$425~cm$^{-1}$, and the 
TO$_{\rm Be-Se}^{\rm Zn}$ mode, at $\sim$475~cm$^{-1}$, show up clearly. 
We recall that these refer to the `long' Be-Se bonds from the Be-rich region 
and the `short' Be-Se ones from the Zn-rich region, respectively. 
We note the parasitical activation of the theoretically forbidden 
LO$_{\rm Be-Se}$ mode around 535~cm$^{-1}$ at $x{\sim}$0.5. This is attributed 
to partial breaking of the selection rules, the result of large 
alloy disorder close to the stoichiometry. All the TO modes become 
blue-shifted when the Be content increases. This is a compression 
effect in response to an overall shrinking of the lattice. We 
note that the linewidths are similar for the two sets of samples, 
both in the Zn-Se and Be-Se ranges, which indicates similar 
crystal qualities. 

Moreover, it is remarkable that the Zn-Se lineshapes from the 
two sets of samples do match almost perfectly after the above 
mentioned normalization, provided a slight overall translation 
along either the ordinate or the abscissa axis is applied. Accordingly 
any difference between the two sets of Be-Se Raman responses 
that can not be suppressed after such overall translation is 
beyond the experimental error. As a matter of fact the agreement 
is not as good in the Be-Se spectral range. The strength ratio 
$R$ between the TO$_{\rm Be-Se}^{\rm Be}$ and the 
TO$_{\rm Be-Se}^{\rm Zn}$ modes is systematically larger for the epilayers 
than for the single crystals, at any $x$ value, testifying thereby an internal 
distortion of the Be-Se signal. This suffices to rule out our 
earlier view that the $R$-anomaly with the ZnBeSe$_2$ epilayer is 
intrinsic (refer to Sec. I). 

A detailed quantitative insight therein is obtained via contour 
modeling of the TO spectra by using a three-oscillator version 
[Zn-Se, (Be-Se)$^{\rm Be}$, (Be-Se)$^{\rm Zn}$] of our phenomenological 
`percolation' model. The Zn-Se and Be-Se input parameters are given elsewhere, 
together with the technical details of the model.$^{11}$ Differently 
from our previous approaches we do not consider any renormalization 
of the BeSe Faust-Henry coefficient in the Zn-rich region. Optimal 
contour modelling of the Raman lineshapes was obtained by taking 
the same phonon damping for the two Be-Se modes. Our principal 
adjustable parameter here is how the available Be-Se oscillator 
strength shares between the two Be-Se modes. Otherwise slight 
frequency adjustment, no larger than $\sim$3~cm$^{-1}$ (refer 
in particular to the low frequency Be-Se mode at $x{\sim}$0.10, 
and to the high frequency Be-Se mode at $x{\sim}$0.50), is applied 
when needed to achieve optimum data modelling. However, no systematic 
trend, that is worth to discuss, could be evidenced with this 
respect. We stress that optimization of the curve fitting procedure 
in the Zn-Se spectral range does not generate any internal distortion 
of the Raman response in the Be-Se spectral range, and \textit{vice 
versa}. The resulting theoretical curves, shifted downwards from 
the data for better clarity, while using the same colour code, 
are shown in Fig.~2, with special emphasis upon the sensitive Be-Se range. 

The $S_{\rm Be-Se}^{\rm Be}$ (circles) and $S_{\rm Be-Se}^{\rm Zn}$
(squares) oscillator strengths (with the conventional notation) 
derived from contour modelling of the TO Raman spectra for the 
whole set of epilayers (black, solid symbols) and the single 
crystals (red, open symbols) are displayed in the main panel 
of Fig.~3. Dotted lines act as guide for eye. The theoretical 
amounts in case of random Be?Zn substitution, i.e.  
$x^2{\cdot}S_{\rm BeSe}^0$ and $x{\cdot}(1-x){\cdot}S_{\rm BeSe}^0$, 
where $S_{\rm BeSe}^0$ is the oscillator strength in pure BeSe 
(refer to Fig. 1), are shown for reference purpose (thin lines). 
While in the single crystals the overall Be-Se oscillator strength 
at the alloy composition $x$ divides consistently with the above expressions, 
as it can be expected for random alloys, in the epilayers the 
amount awarded to the TO$_{\rm Be-Se}^{\rm Be}$ (TO$_{\rm Be-Se}^{\rm Zn}$)
 mode is systematically larger (smaller) than expected. This 
is consistent with CPSO, i.e. over (sub) representation of the 
`long' (`short') Be-Se bonds from the Be-rich (Zn-rich) region 
as schematically indicated by the curved arrow at the bottom 
of Fig.~2 (refer also to the upper curved arrow in Fig.~1), with 
concomitant impact on the local strain energy as discussed above. 
The reverse trend would enlarge the overall contrast between 
the Zn-Se and Be-Se bond lengths in the alloy, and is thereby 
forbidden by SO in principle. 

\begin{figure}[t]
\centerline{%
\includegraphics[height=0.7\textwidth,clip=true,angle=270]{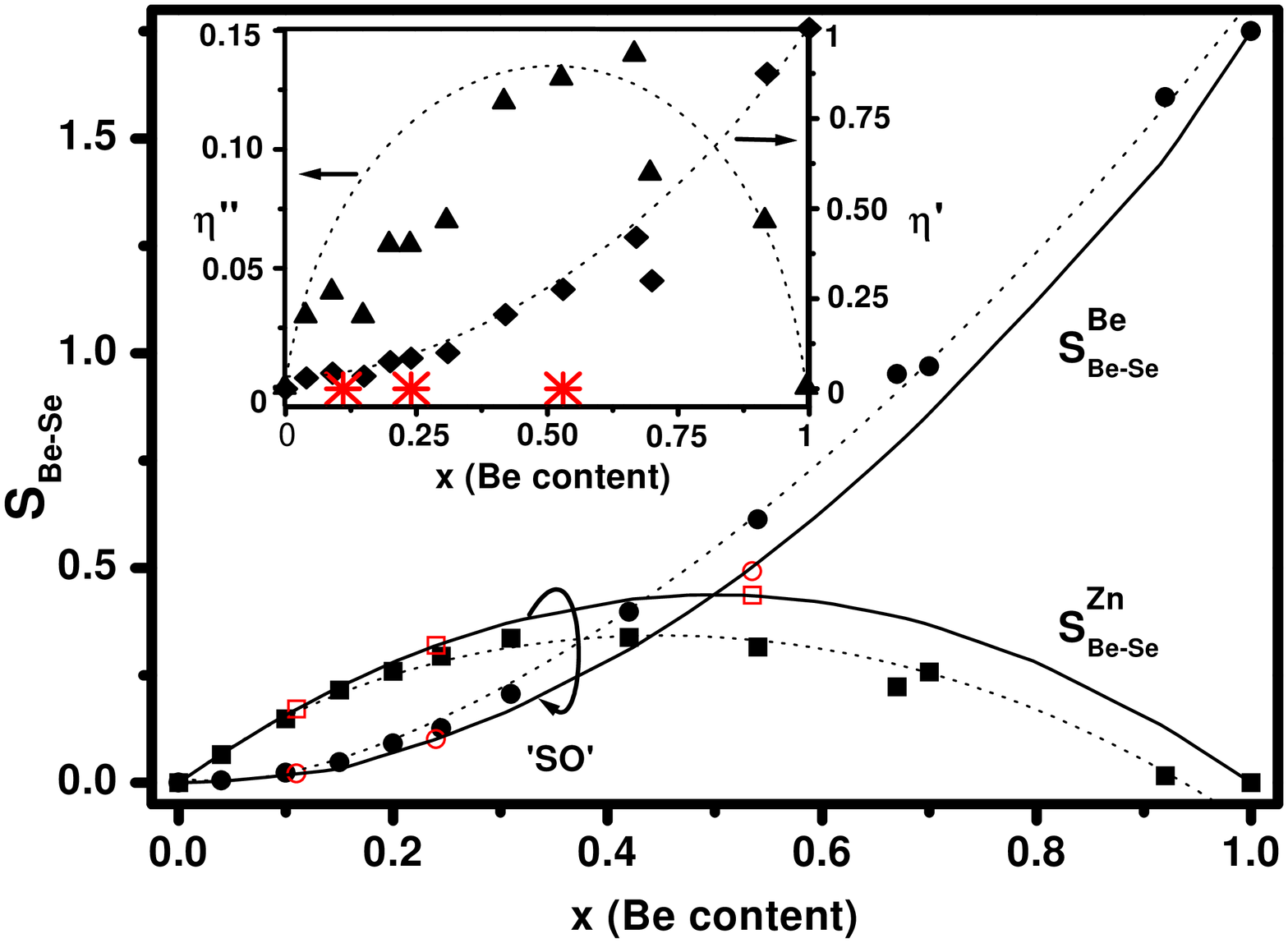}
}
\caption{%
(color online): $S_{\rm Be-Se}^{\rm Be}$ (circles) and
$S_{\rm Be-Se}^{\rm Zn}$ (squares) oscillator strengths in the single crystals
(red, open symbols) and the epilayers (black, plain symbols), as derived
from contour modeling of the multi-mode TO Raman spectra (refer to Fig. 2)
in the Be-Se spectral range. The theoretical values in the random case are
added (thin lines), for reference purpose. The deviation with respect to
the random case is taken as an effect of SO, as schematically indicated
by the curved arrow (refer also to the upper curved arrow in Fig. 1). The
$\eta'$ (diamonds) and $\eta''$ (triangles) values obtained
with the epilayers (black) are shown in the inset. The corresponding values
for the single crystals are also shown (red stars, same symbol for the two
data sets), for comparison. In the body of the figure as well as in the inset
the dotted curves are guidelines for eye.}
\end{figure}

Interestingly the Zn-Se signal from the ZnBeSe$_2$ epilayer (refer 
to the spectrum in black at the bottom of Fig.~2), which reveals 
the most spectacular effect in the Be-Se range, is slightly blue-shifted 
with respect to the single-crystal reference. This is exactly 
the trend expected in case of CPSO, i.e. the result of reinforcement 
of the `short' Zn-Se species, vibrating at high frequency, to 
the detriment of the `long' species, vibrating at low frequency 
(refer to the lower curved arrow in Fig.~1). Now, care must be 
taken that the discrete Zn-Se Raman signal is corrupted by a 
parasitic Fano interference with a disorder-activated acoustical 
continuum at large $x$ values, as evidenced by the characteristic 
antiresonance at $\sim$200~cm$^{-1}$ (Ref.~11). This does 
not allow us to be fully conclusive. Reliable insight into the 
actual dependence of the vibrational properties in the Zn-Se 
spectral range versus ordering is obtained by theoretical means in Sec.~IV.

We denote as $\eta'$ and $\eta''=(1-x)\!\cdot\!{\eta'}$ the fraction of `short' 
Be-Se bonds that have turned `long' due to SO, and their corresponding fraction
with respect to the total number of bonds in the crystal, respectively. 
$\eta'$ is given by the relative decrease of $S_{\rm Be-Se}^{\rm Zn}$
with respect to the random case $(\eta'=\eta''=0)$, i.e. 
$\eta'=(S_{\rm Be-Se}^{\rm Zn} - 
S_{\rm Be-Se}^{\rm Zn,r})/S_{\rm Be-Se}^{\rm Zn,r}$, where the additional 
superscript `r' refers to the random case. $\eta''$ follows directly. 
In practice $\eta'$ is a measure of the transfer of Be-Se oscillator strength 
from the higher frequency mode to the lower frequency one, as a deviation 
with respect to the nominal partition of Be-Se oscillator strength 
in the random case (refer to Fig.~1). The $\eta'$ (diamonds) and 
$\eta''$ (triangles) values obtained with the epilayers are shown in 
the inset of Fig.~3. The $\eta'$ vs. $x$ curve corresponds to a smooth 
parabolic-like divergence, which indicates that the larger $x$, the more 
easily the `short' Be-Se bonds convert into the `long' ones. By comparison, 
$\eta'=\eta''=0$ for the single crystals, at any $x$ value 
(refer to the red stars in the inset of Fig.~3). From the $\eta''$
vs. $x$ curve, less than $\sim$15\% of the bonds would be concerned 
with CPSO in the epilayers. Remarkably, CPSO would be activated 
from the dilute limits. It gets maximum at $x{\sim}$0.5 
($\eta''{\sim}$0.15). At this limit the Be-Se and Zn-Se bonds are 
equally represented in case of random Be substitution to Zn, 
as are the `short' and `long' variants for each bond species, thus 
creating a situation where four different bond lengths coexist 
in the alloy with identical populations. This results in a maximum 
local strain energy in the crystal, hence making more appealing 
an option to reduce the strain through the development of SO. 

It is important to note that the ($\eta'$, $\eta''$) vs. $x$ curves 
do not exhibit any singularity around the critical $x$ values of 
$\sim$0.25 and $\sim$0.6 (refer to Sec. II). This indicates that these curves 
should not be discussed in relation to the substrate-induced residual tensile 
strain in the epilayers. This legitimates \textit{a posteriori} our above 
discussion of the observed transfer of Be-Se oscillator strength in terms 
of an intrinsic process, i.e. along the broad lines developed at the 
beginning of this section (refer to subsection A). 

The possibility of the CPSO in the Zn$_{1-x}$Be$_x$Se epilayers gets 
additional support from the bowing of the optical bandgap ($E_g$) 
observed for our two sample sets, and reported in the literature. 
Indeed it is well known that the CPSO reduces $E_g$.$^6$ As a 
matter of fact, the existing data indicate that the free exciton 
recombination line is blue-shifted at the constant rate of 24.10~meV 
per \%Be throughout the sequence of our single crystals ($x$=0.4, 
$T$=40~K),$^31$ and at the smaller rate of 23.00~meV per \%Be with 
our epilayers ($x$=0.6, $T$=10~K).$^{32}$ Actually this is consistent 
with our view that the epilayers exhibit the CPSO to some extent, 
as inferred from the Raman data. Incidentally the extra bowing 
of the optical bandgap in the case of the epilayers can not be 
attributed to the residual tensile strain due to the lattice 
mismatch at the ZnBeSe/GaAs interface. Indeed our epilayers are 
fully relaxed above $x{\sim}$0.25 (refer to Sec. II), and Tourni\'{e} 
\textit{et al.}$^{32}$ did not detect any singularity in the dependence of 
the optical bandgap on $x$ around this critical alloy composition. 
Note that from 10~K to 40~K the excitonic line of our ZnSe and 
Zn$_{0.59}$Be$_{0.41}$Se epilayers are red-shifted by $\sim$50~meV$^{33}$ 
and $\sim$35~meV (see Fig.~21, p.~96 in Ref.~25), respectively, 
so that at 40~K the epilayer rate is renormalized to 23.35~meV 
per \%Be, i.e. closer to the rate found for the single crystals. 
Actually, the difference is small. By comparison, the rate for 
perfectly-ordered Zn$_{1-x}$Be$_x$Se is $\sim$11.60~meV per \%Be 
($x$=0.5, $T$=0~K), as determined from recent first-principles calculations 
performed by Tsai \textit{et al.} with fully relaxed supercells.$^{34}$ 
If we focus on ZnBeSe$_2$, the $E_g$-difference between our single 
crystal and our epilayer would be $\sim$50~meV, representing only $\sim$8\% 
of the total $E_g$-reduction from random (our single crystal) to 
perfectly-ordered ZnBeSe$_2$. When brought back to the theoretical $E_g$ vs. 
$\eta$ curve now available for the reference InGaP$_2$ system,$^{10}$ 
this is just beyond the $E_g$-fluctuation at $\eta$ $\sim$0, corresponding 
to $\eta{\sim}$0.1. While such $\eta$ value is generally consistent with 
a small amount of CPSO in the epilayers ($\eta''$=0.15, see above), as expected
(refer to Sec.~I), we must admit that the observed trend in the bandgap 
can not yet serve as evidence for ordering by itself. Indeed in 
Zn$_{1-x}$Be$_x$Se, where the ordering is by far less pronounced than 
in Ga$_{1-x}$In$_x$P, the lowering of the optical bandgap, especially as small 
as that reported here, can be attributed to many reasons besides the SO.

Certainly, the technique of choice to detect the CPSO would be 
diffraction of electrons. The CPSO would manifest itself via 
the emergence of additional features in the electron diffraction 
pattern, corresponding to an additional periodicity due to the 
alternation of Ga-rich and In-rich monolayers along the [111] 
direction, on top of the basic periodicity from the lattice. 
However, given the experience concerning GaInP$_2$, it seems improbable 
that CPSO could be revealed by electron diffraction for $\eta$
values lower than $\sim$0.2, as is apparently the present case. 

More generally, $\eta{\sim}$0.2 seems to be the limit for proper detection 
of the CPSO by any of the traditional experimental means. A possible 
alternative to detect the CPSO below this limit is our suggestion 
to measure a deviation in the ratio of strength between the two 
TO Raman lines that refer to the same bond species, with respect 
to the nominal ratio for random alloys (as derived from the fractions 
given within square brackets in Fig.~1). In particular at the 
stoichiometry, where the CPSO has just been argued to be most likely to occur, 
this must be a quite sensitive method because the two modes at 
$\eta$=0 are expected to show up with similar intensities, so that 
any fluctuation in the above ratio should be easily detectable. 
Basically there is no intrinsic limit of detection here. We recall 
that the puzzling interplay between the strengths of the two 
Ga-P Raman modes (TO) of the reference GaInP$_2$ alloy when ordering 
increases was successfully explained on this very basis recently.$^{13}$ 
This method seems to be sensitive in the dilute limit also. Indeed we have 
identified earlier in N-dilute InGaAsN such interplay between those two 
Ga-N modes that come from the N-rich and the (Ga,In)-rich regions,$^{35}$ 
apparently the result of CuPt type ordering.$^{21}$

We admit as an intermediate summary that the CPSO in our ZnBeSe 
epilayers, as inferred from the Raman data, could not be so far 
independently confirmed by other experimental means, due to a 
presumably too low degree of ordering which falls below the detection 
threshold. However, our Raman approach is validated by an analogy 
with GaInP$_2$. Independent support comes from the theoretical 
side, in the form of first-principles phonon/bond length calculations 
discussed below. 

\section{FIRST-PRINCIPLES CALCULATIONS 
(Z\lowercase{n}B\lowercase{e}S\lowercase{e}$_2$): AN 
INSIGHT AT THE MICROSCOPIC SCALE }
Here we address the issue ($iii$) introduced in Sec.~I, about the 
nature of the microscopic mechanism that should hinder CPSO in 
ABC$_2$ alloys beyond $\eta{\sim}$0.5. For direct insight into this key issue, 
we performed first-principles calculations of the dependence of the bond length 
and phonon properties in ZnBeSe$_2$ on the degree of ordering, 
taken from the random limit ($\eta$=0) up to the formation of the CuPt-type 
superstructure ($\eta$=1). 

\begin{figure}[bt]
\includegraphics[width=0.75\textwidth,clip=true,angle=0]{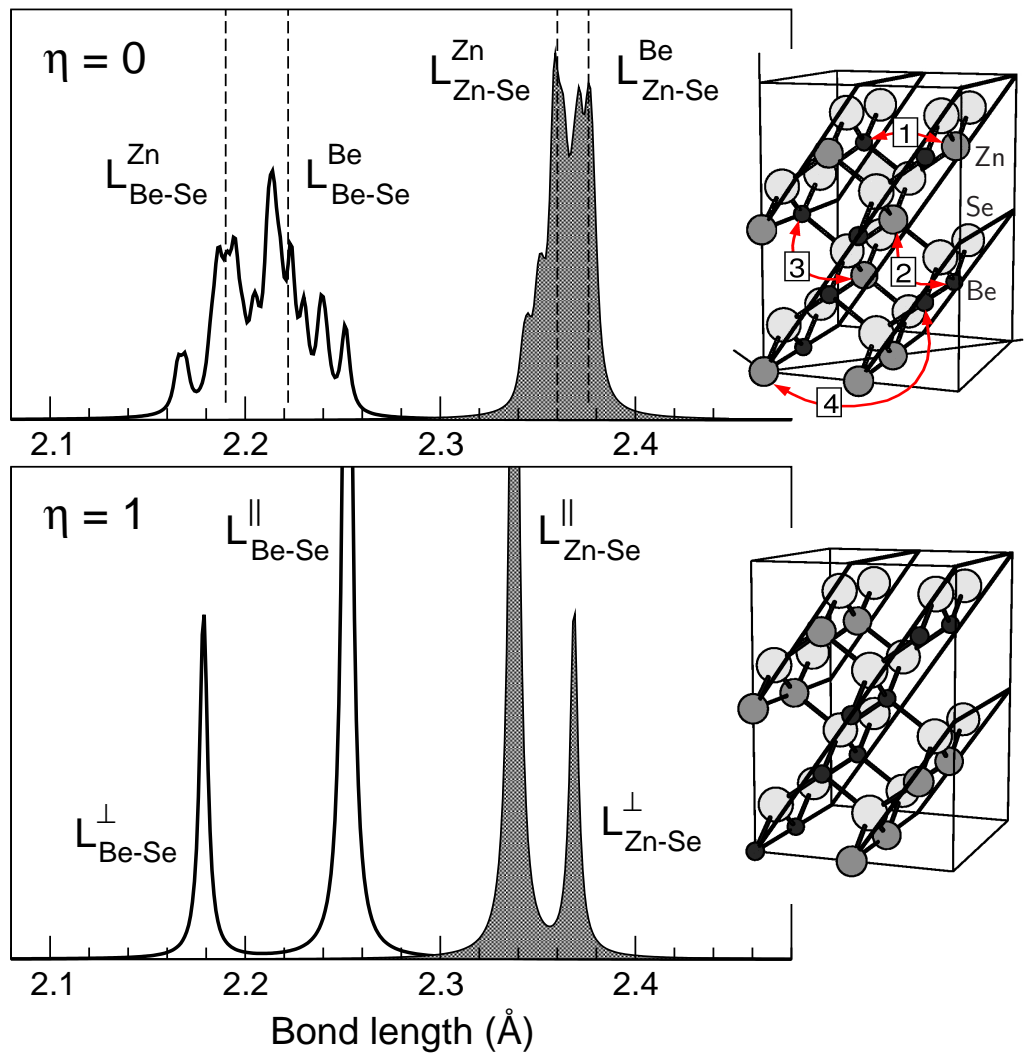}
\caption{%
Overall Zn-Se and Be-Se bond length distributions, broadened
with halfwidth 0.0025 {\AA}, of random ZnBeSe$_2$ alloy ($\eta{=}0$, top panel)
and of the full CuPt-type ordered ZnBeSe$_{2}$ alloy ($\eta{=}1$,
bottom panel), as derived with the supercells shown on the right.
The labelling of the individual features comes from the `percolation' 
(regime 1) and `superlattice' (regime 2) terminologies,
respectively. The finite $\eta$ values of 0.25, 0.5, 0.75 and 1 can be
achieved by cumulating Zn-Be pair exchanges from 1 to 4, as indicated
by the arrows in the top supercell.}
\end{figure}

We used Zn$_8$Be$_8$Se$_{16}$ 32-atom supercells, spanned by the vectors 
$[1\bar{1}0]$, $[110]$ and $[002]$ in units of the lattice constant of (cubic) 
unit cell of the zincblende structure. Se atoms retain the anion positions, 
whereas cation sites are occupied by Zn and Be in such a way as to achieve 
a desired value of $\eta$. The $\eta$=0 supercell (shown in the top panel 
of Fig.~4) assures equal distribution of Zn and Be over (111) cationic planes 
(in the original cubic zincblende setting), irrespectively of a possible 
orientation of the (111) plane family. The supercells corresponding to 
$\eta$=0.25, 0.50, 0.75 and 1.00 (for the latter, refer to the bottom 
panel of Fig.~4) are then constructed by gradually interchanging 
one (Be,Zn) pair after another, between consecutive cationic 
planes. The successive swaps are labelled as 1 to 4 in the starting 
supercell shown at the top of Fig.~4. A deviation of $\eta$ from zero 
arbitrarily fixes a family of (111) planes, as is also shown in the supercell 
drawings. The construction of supercell for a given $\eta$ value is 
an otherwise ambiguous procedure. A full statistical analysis over all 
possible supercell choices compatible with a given $\eta$ value would be 
too computationally demanding. Our aim here is to offer representative 
supercells for different $\eta$ values, and to discuss the trends 
in the distribution of bond length and phonon frequencies, as the pattern 
of cation-anion connectivity varies with $\eta$ . 

\begin{figure}[t]
\includegraphics[width=0.75\textwidth,clip=true,angle=0]{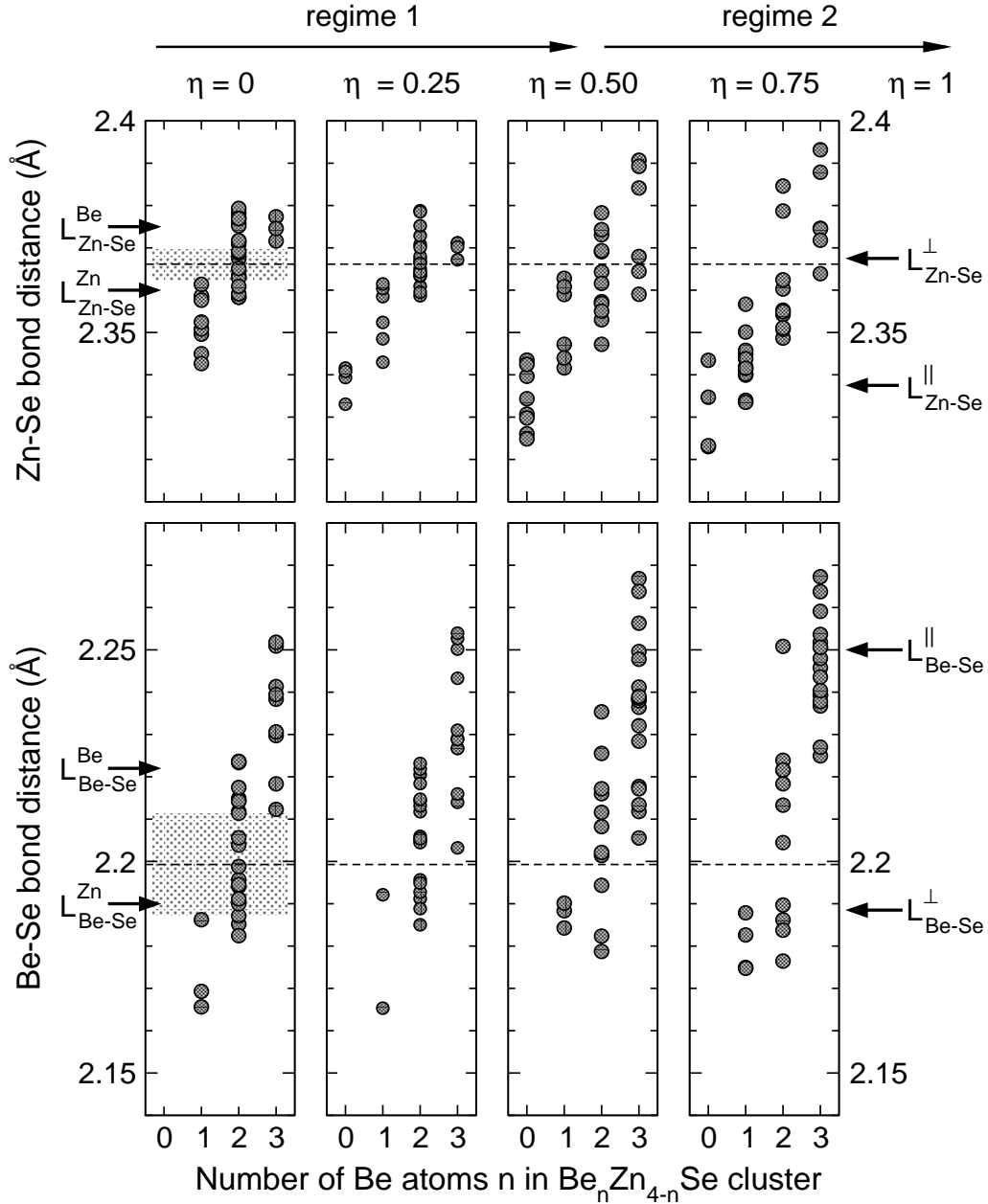}
\caption{%
Zn-Se (top) and Be-Se (bottom) bond length distributions
for Se-centered Zn$_{4-n}$Be$_n$ ($n$=0--4) tetrahedron clusters in
the supercells with different $\eta$ values (refer to Fig.~4).
The Zn-Se and Be-Se 1-bond$\rightarrow$2-mode bond length situations 
at the two limiting $\eta$ values are schematically indicated by arrows, 
for reference purpose.
The Zn-Se and Be-Se bond lengths at the middle of the gaps (shaded areas) 
between the data sets related to $n$=1 and $n$=3 at $\eta$=0 are indicated 
by dashed lines, for qualitative insight upon the evolutions of the Zn-Se 
and Be-Se distributions of bond lengths in the Be-rich and Zn-rich regions 
when ordering ($\eta$) increases. Basically regime 1, corresponding to clear 
convergence of the average Zn-Se and Be-Se bond lengths, is relayed by regime
2, corresponding to a \textit{status quo}, as indicated at the top
of the figure.}
\end{figure}

For each supercell chosen, we allow a full unconstrained optimization 
of structure (lattice constants and internal parameters) and 
obtain a certain scattering of (64 in total) bond lengths, which 
are shown (with artificially introduced broadening) in the body 
of Fig.~4, for the extreme cases $\eta$=0 (top panel) and $\eta$=1 
(bottom panel). The perfectly CuPt-ordered case $\eta$=1 is especially simple 
(see below). The distribution of bond lengths for $\eta$=0 and other 
$\eta$ values is more obscure and shown in Fig.~5, for the sake of 
more detailed analysis, split into several groups, according 
to the number $n$ ($n$=0-4) of Be atoms in the Se-centred Zn$_{4-n}$Be$_n$ 
host tetrahedron clusters. 

With the structure fully optimized for each supercell, we calculate 
the spectrum of phonon frequencies of the zone-centre of the 
supercell, by the frozen phonon technique. As we are particularly 
interested in comparison with the Raman spectra which probe the 
zone-centre phonons of the underlying primitive cell, we recover 
the density of states of transversal phonons (TO-DOS) by projecting 
each vibration mode in the supercell onto the uniform translation 
of the primitive zincblende cell. In doing this, we follow an 
approach earlier used in Ref.~11, specifically Eq.~(4) therein. 
The resulting TO-DOS curves for the different $\eta$ values are shown 
in Fig.~6.

\begin{figure}[t]
\includegraphics[width=0.6\textwidth,clip=true,angle=0]{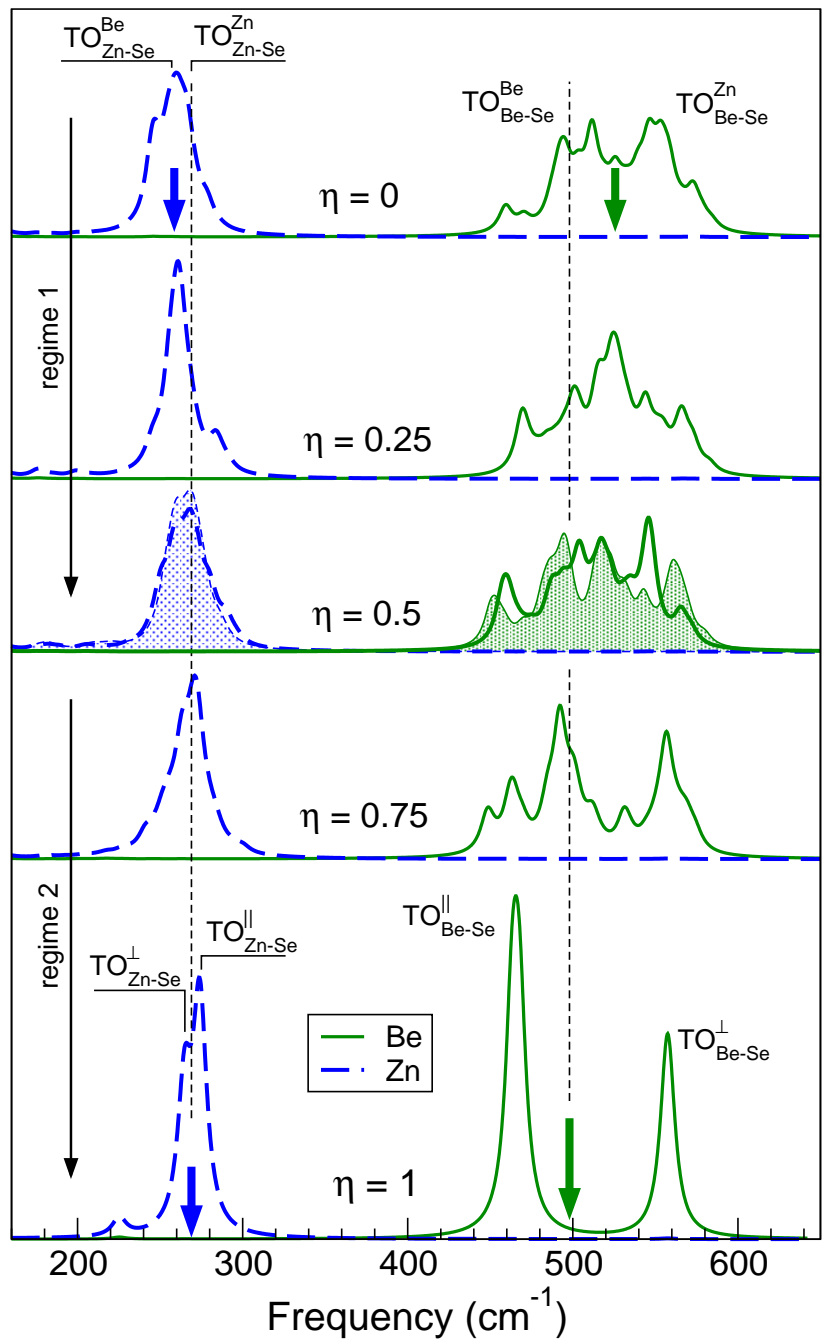}
\caption{%
Evolution of the TO-DOS of the Zn (dashed lines) and Be
(solid lines) atoms when ordering ($\eta$) increases (in the row of
supercells explained in Fig.~4). The arrows indicate the barycenters
of the TO-DOS at the limiting $\eta$ values. The vertical dotted lines mark
the barycenters at $\eta=1$, for reference purpose. Two different regimes 
labelled as 1 and 2, corresponding to regimes 1 and 2 in Fig.~5, can be 
distinguished on two sides of $\eta{\sim}0.5$. At this limit the TO-DOS 
obtained by inverting all Zn and Be atoms in the original supercell, thereby 
leaving the $\eta$ value unchanged, is superimposed (dashed areas) 
to the original curve,  for comparison.}
\end{figure}

\subsection{Phonon / bond length situation in the random 
($\eta$=0) and CuPt ordered ($\eta$=1) crystals}
First we discuss the phonon/bond length behaviours at the two extreme 
$\eta$ values, for reference purpose. 

At $\eta$=1 ZnBeSe$_2$ consists of a [111]-oriented ZnSe/BeSe monolayer 
superlattice (refer to the supercell at the bottom of Fig.~4). 
Lattice-matching at the interface is achieved via in-plane hydrostatic 
tension (compression) and out-of-plane compression (tension) 
of the BeSe (ZnSe) layer. Accordingly the bond lengths, on the 
average well separated into short Be-Se and long Zn-Se bond 
species, do further split into three identical in-(111)-plane 
ones (longer than average for Be-Se, shorter than average for 
Zn-Se) and a single out-of-(111)-plane one. Our notation is 

$L_{\rm (Zn,Be)-Se}^{\parallel, \perp}$
 (refer to the bottom panel of Fig.~4, and also to the right 
side of Fig.~5), where the subscript refers to the bond species, 
and superscripts $\parallel$ and $\perp$
 represent in-plane and out-of-plane bonds, respectively. The 
TO-DOS features in Fig.~6 are correspondingly labelled, just 
replacing L by TO.

Now we turn to $\eta$=0. Here the TO-DOS (top curve of Fig.~6) exhibits, 
within each of the Zn-Se ($\sim$270~cm$^{-1}$) and Be-Se 
($\sim$525~cm$^{-1}$) spectral ranges, two distinct phonon modes of similar 
strengths, as could be ideally expected from the `percolation' picture (refer 
to the relative strengths of the individual modes, as indicated 
within square brackets in Fig.~1). For unambiguous insight in 
the Zn-Se spectral range, refer to the discussion of Fig.~7 (left 
panel) below. Therefore for labelling the modes we retain the 
earlier introduced terminology of the percolation picture, i.e. 
TO$_{\rm (Zn,Be)-Se}^{\rm Be,Zn}$. Similar labelling, with TO replaced by L, 
is used to mark the corresponding dominant lines in the Be-Se 
and Zn-Se bond length distributions (refer to the top panel of Fig.~4, 
and also to the left side of Fig.~5). Incidentally, the present calculations 
provide the first evidence for 1-bond$\rightarrow$2-mode (TO) behaviour 
at the stoichiometry in Zn$_{1-x}$Be$_x$Se, in both the Zn-VI and Be-VI 
spectral ranges. 

\begin{figure}[t]
\includegraphics[width=0.77\textwidth,clip=true,angle=0]{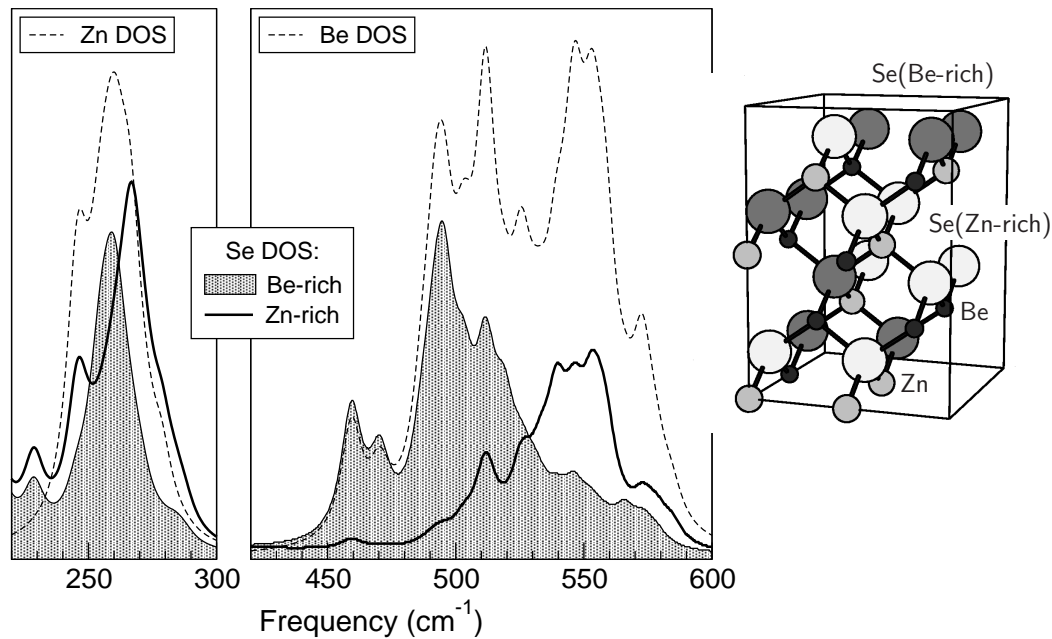}
\caption{%
Partition of the Se atoms in the supercell at $\eta{=}0$ (refer to Fig.~4,
top supercell) into those from the Be-rich (dark) and Zn-rich (light) regions, 
as inferred from the individual Se TO-DOS. The accordingly partitioned 
Se TO-DOS is compared to those of Zn and Be (dashed curves, 
arbitrarily scaled).}
\end{figure}

Let us compare briefly the obtained TO-DOS curve at $\eta$=0 with 
the corresponding Raman data (refer to the spectra at the bottom of Fig.~2), 
to check the validity of our first-principles calculations. We note that 
the calculated and experimental Be-Se phonon splittings are similar, 
i.e. of $\sim$45 cm$^{-1}$, even though the TO-DOS curve is globally 
blue-shifted by about $\sim$50 cm$^{-1}$ with respect to the Raman spectra, 
an effect of slightly too compressed lattice, due to a well-known overbinding 
of the local density approximation. The same overall blue-shift was 
observed in our earlier study at the Be-Se bond percolation threshold,
$\sim$19\% of Be.$^{12}$ Further on, the corresponding difference 
in bond length, i.e. $\delta{\sim}$1.35\% (refer to the dotted 
lines in the top panel of Fig.~4), is similar to that earlier 
estimated at the latter limit.$^{12}$ This is consistent with quasi-invariance 
of the Be-Se phonon splitting under change of the alloy composition, 
as can be inferred from the Raman data (refer to Fig.~1). On 
the Zn-Se side the calculated difference in bond length reduces to
$\delta{\sim}5{\textperthousand}$ (see Fig.~4, top panel), with 
concomitant impact on the phonon splitting, estimated as $\sim$10~cm$^{-1}$ 
(see Fig.~7, left panel, discussed below). No direct comparison 
with the experimental ZnBeSe$_2$ data is possible here, because 
the Zn-Se Raman signal is strongly distorted by a parasitic 
Fano-interference, as already mentioned. One can turn to the 
Zn-Te Raman response from the similar Zn$_{1-x}$Be$_x$Te alloy (the Be-VI 
phonon splitting at the stoichiometry is $\sim$25~cm$^{-1}$ here, see Fig.~1 
in Ref.~16). A direct insight at the stoichiometry is hindered because 
the Zn-Te signal is too weak to be detected,$^{36}$ but we argued above that 
the estimate at the Be-Te bond percolation threshold could serve as 
a substitute. The observed splitting of $\sim$8~cm$^{-1}$ (Ref.~36) 
is in reasonable agreement with the present calculation estimate for 
the Zn-Se species. 

Now we discuss the bond length aspect. What emerges from Fig.~5 is that 
the longer (shorter) Be-Se and Zn-Se bonds come from Se-centred tetrahedra 
with a large (small) number of Be atoms at the vertices, i.e. 2-3 (1-2). 
Note that a similar trend was earlier observed by Silverman et al.$^{37}$ 
in their first-principles calculations dedicated to GaInP. Introducing 
a link to the earlier proposed terminology of the `percolation' picture, 
we shall refer to such `long' (`short') bonds as those belonging to the Be-rich 
(Zn-rich) region. The important point here is that none of the Be-rich 
and Zn-rich regions reduces to a single Se-centred tetrahedron 
species. This implies that the 1-bond$\rightarrow$2-mode (TO,L) behaviour 
can not be discussed in terms of a mere tetrahedron problem. 

More precisely, what is certain is that the Se atoms surrounded 
by 1 and 3 Be atoms all belong to the Zn-rich and Be-rich regions, 
respectively. We take advantage of this clear discrimination 
and mark the Zn-Se and Be-Se bond lengths just in-between these 
two data sets (refer to the dotted lines in the middle of the 
shaded areas at $\eta$=0 in Fig.~5), for straightforward -- qualitative -- 
insight upon the evolutions of the Zn-Se and Be-Se distributions of 
bond lengths in the Zn-rich and Be-rich regions as the degree of CuPt ordering 
increases. Among the Se atoms with 2 Be atoms as first neighbours, 
some belong to the Zn-rich region and others to the Be-rich one, 
but this discrimination is less straightforward. A possible partition 
criterion follows from examining the individual atom-resolved TO-DOS for each 
Se atom (not shown). A given Se atom is classified as part of the Be-rich 
(Zn-rich) region if the barycentre of its individual TO-DOS is located on 
the low (high) frequency side, both in the Be-Se and Zn-Se spectral ranges. 
Obviously this criterion is not free from ambiguity, because some Se atoms 
vibrate at lower as well as at higher frequencies throughout each of the Zn-Se 
and Be-Se spectral ranges. Our separation of the Se atoms into those belonging 
to the Be-rich region and to the Zn-rich one in our $\eta$=0 supercell is shown
in Fig.~7, along with the partial TO-DOS corresponding to these two groups 
of Se atoms. As expected, the partial TO-DOS of the Se atoms from the Be-rich 
and Zn-rich regions mimic fairly well the individual 
$TO_{\rm (Zn,Be)-Se}^{\rm Be}$ and $TO_{\rm (Zn,Be)-Se}^{\rm Zn}$
features from the TO-DOS related to the Zn and Be atom species 
(shown by dotted lines), respectively, in the corresponding spectral ranges.

From the supercell shown in Fig.~7 it appears that the two sub-continua 
formed by the Se atoms from the Be-rich and Zn-rich regions are 
finely interlaced, as predicted by the `percolation' picture.$^{11}$
One crucial question is whether there exists any specific feature 
in the topologies of the Se atoms from the Be-rich and Zn-rich 
regions that can be turned into a `microscopic rule' for proper 
identification of the atoms/bonds from one or the other region. 
Unfortunately, a careful examination of the two Se-topologies 
does not reveal any straightforward `rule'. Probably the degree 
of connectivity between the individual bonds of like species 
should play a role, although a proper criterion in relation to 
connectivity could not be derived in our case. Our present view 
is that the topologies of the Be-rich and Zn-rich regions should 
be more suitably defined at the mesoscopic scale, by using some 
proper fractal exponent. Further discussion of this aspect falls 
beyond the scope of this work. Generally, the whole of this is 
consistent with our description of the 1-bond$\rightarrow$2-mode (TO,L) 
behaviour in terms of a percolation-based phenomenon. 

At this stage we can state that our first-principles calculations 
successfully account for a 1-bond$\rightarrow$2-mode (TO,L) 
type behaviour not only in the perfectly-ordered CuPt-type alloy 
($\eta$=1), which is rather obvious, but also in the random alloy 
($\eta$=0), as observed in the Raman spectra, which was not so 
obvious. 

\subsection{Phonon / bond length situation in its dependence on the degree 
of ordering ($0<{\eta}<1$)}
Now we discuss the full $\eta$-dependence of the bond length data (Fig.~5) 
and of the TO-DOS (Fig.~6) under the above angle. Two regimes, different 
in nature, can be distinguished: regime 1, directly issued from 
the peculiar 1-bond$\rightarrow$2-mode (TO,L) percolation-type situation 
at $\eta$=0; and regime 2, just the progressive building up of 
the typical 1-bond$\rightarrow$2-mode (TO,L) superlattice-type behaviour 
at $\eta$=1. For more clarity in the discussion below, we mark by arrows 
the barycentres of the Zn-Se and Be-Se TO-DOS at the extreme $\eta$ values 
in Fig.~6.

In regime 1 (0<${\eta}{\leq}$0.5, typically) the starting ($\eta$=0) 
1-bond$\rightarrow$2-mode (TO,L) Zn-Se and Be-Se TO-DOS seem to turn 
progressively 1-bond$\rightarrow$1-mode (TO,L) in character, simultaneously 
re-centring upon the individual Zn-Se and Be-Se modes that face 
each other at $\eta$=0, i.e. $TO_{\rm Zn-Se}^{\rm Zn}$ and 
$TO_{\rm Be-Se}^{\rm Be}$. This indicates that the transfer of Be-Se 
oscillator strength detected in the Raman spectra of our epilayers (Sec.~III-B)
is actually consistent with CPSO. The observed transfer is small 
in the present case, as already discussed, signing the emergence 
of CPSO only, i.e. the onset of regime 1.

The end of regime 1, is apparently reached at $\eta{\sim}$0.5 (refer to 
the middle curve in Fig.~6) -- recall our issue ($iii$). Indeed at this limit 
both the Be-Se and Zn-Se phonon responses seem of a 1-bond$\rightarrow$1-mode 
(TO) type, although the features are rather broad. More precisely the Be-Se 
and Zn-Se responses apparently reduce to the only $TO_{\rm Zn-Se}^{\rm Zn}$
and $TO_{\rm Be-Se}^{\rm Be}$ modes, respectively, that have attracted 
the quasi-totality of the oscillator strength in each spectral range. 
Incidentally we have checked that the TO-DOS obtained by inverting Zn and 
Be atoms in the $\eta{\sim}$0.5 supercell, thereby leaving the $\eta$
value unchanged, is very similar to the original one (refer to the shaded 
curves at $\eta{\sim}$0.5 in Fig.~6). This confirms that the 
`full condensation' at $\eta{\sim}$0.5 is not fortuitous, the result of 
a favourable atom arrangement, but actually intrinsic, fixed by the 
$\eta$ value only. The whole of this is consistent with the bond length 
data in Fig.~5. At $\eta$=0.5, only 6 Be-Se (8 Zn-Se) bonds out of 32 remain 
`short' (`long'), i.e. still belong to the Zn-rich (Be-rich) region 
(compare the data on each side of the reference dotted lines in Fig.~5), if 
using the terminology of the `percolation' picture. A crucial point is that 
at $\eta{\sim}$0.5 the local bonding between the anionic and cationic 
(111) planes still has a (Zn-Se, Be-Se)-mixed character (see the atom 
arrangement in the $\eta{\sim}$0.5 supercell), as in a random alloy.  Regime 1 
is schematically represented by the antagonist curved arrows in Fig.~1.

With further Zn$\leftrightarrow$Be exchange beyond $\eta{\sim}$0.5, 
the (Zn-Se, Be-Se)-mixed character of the local bonding between the cationic 
and anionic planes is lost, due to the formation of ZnSe/BeSe superlattice-like
micro-domains here and there (see the atom arrangement in the 
$\eta$=0.75 supercell). The alloy enters the so-called regime 2, in 
which the Be-Se and Zn-Se TO-DOS, and also the bond length distributions, 
turn bi-modal again, the result of the difference between the 
in-plane and out-of-plane bond lengths within these micro-domains 
(refer to the discussion at $\eta$=1). The trend is emphasized progressively 
until perfect CuPt-ordering. The key point is that, in contrast with regime 
1, regime 2 does not seem much rewarding with respect to further 
minimization of the local strain energy in the crystal. This 
is clear especially from Fig.~5, where ordering in regime 2 leaves 
the overall difference between the Zn-Se and Be-Se bond lengths 
quasi-unchanged with respect to the $\eta{\sim}$0.5 situation. The same 
conclusion can be derived also from quasi-invariance of the barycentres 
of the Zn-Se and Be-Se TO-DOS beyond $\eta{\sim}$0.5 (compare the locations 
of the dotted lines and of the arrows at $\eta$=1 in Fig.~6). 

Basically it seems that the cause for further SO is suppressed 
in regime 2, that is the answer to issue ($iii$) raised in Sec.~I. What 
emerges is that the end of regime 1 should be an intrinsic limit to CPSO. 
It is important to notice that with the traditional 
1-bond$\rightarrow$1-mode (TO,L) description at $\eta$=0, the regime 1 is 
totally suppressed, and along with it the notion of an intrinsic limit to CPSO,
being an essential element of this regime, disappears as well. 

If we recollect the reference GaInP$_2$ system, it is very interesting 
to note that $\eta{\sim}$0.5 corresponds to a singularity in the evolution of 
the bond length distribution versus the order parameter,$^{10}$ 
as already discussed. In the latter work the bonds were discriminated 
\textit{a priori} according to whether they were along the ordering direction 
or along the lateral direction, at any $\eta$ value, leading automatically 
to a discussion of the observed behaviours in terms of superlattice-type 
effects only, corresponding to our regime 2 essentially. However, the nature 
of the singularity at $\eta{\sim}$0.5 could not be explained on this basis, 
neither why this critical $\eta$ value just corresponds to the strongest 
ordered GaInP$_2$ samples currently available. Our present view is that 
the singularity in the bond length distribution of GaInP$_2$ parallels 
the singularity in the phonon mode behaviour of ZnBeSe$_2$, and should be 
discussed in terms of a transition from regime 1 to regime 2, with, at 
the end, the proper identification of $\eta{\sim}$0.5 as an intrinsic limit 
to SO in GaInP$_2$. As a matter of fact we recall that GaInP$_2$ was shown 
to exhibit the same phonon mode behaviour as ZnBeSe$_2$ at $\eta$=0.$^{13}$
Therefore this alloy should go through regime 1 before entering regime 2 
when ordering increases. 

In summary our view is that in ABC$_2$ alloys, the A$\leftrightarrow$B exchange 
between consecutive substituting planes is advantageous for minimizing 
the local strain energy until AC/BC superlattice-like micro-domains 
are formed, and this eventually hinders a further increase of 
CPSO. This end is reached at $\eta{\sim}$0.5, the junction of regimes 1 and 2. 
At this limit there is a pure 1-bond$\rightarrow$1-mode behaviour in the TO 
Raman spectra. Apart from this special situation the phonon mode behaviour is 
of a 1-bond$\rightarrow$2-mode type, whereby this behaviour is of different 
nature on two different sides of $\eta{\sim}$0.5, i.e. of a percolation-type
at lower $\eta$ value and of superlattice-type at higher $\eta$ value.

\section{CONCLUSION}
We tackle the key issue of CPSO by assuming a 1-bond$\rightarrow$2-mode 
(TO,L) behaviour for the random A$_{1-x}$B$_x$C zincblende mixed 
crystals as accounted for by our so-called `percolation' picture 
at the mesoscopic scale, which markedly deviates from the traditionally 
admitted 1-bond$\rightarrow$1-mode (TO,L) description at the macroscopic 
scale. In particular, on this novel basis it becomes possible 
to identify an \textit{intrinsic} driving force behind the spontaneous 
local segregation of the A and B species within different, intercalated, 
series of substituting planes, with interleaving planes of the 
other unperturbed C species. As a case study we discuss the phonon 
mode behaviour of the `model' Zn$_{1-x}$Be$_x$Se system, that exhibits 
a clear 1-bond$\rightarrow$2-mode behaviour in the Be-Se spectral range, 
from both the experimental and theoretical sides. 

On the experimental side, we discuss the different 1-bond$\rightarrow$2-mode 
TO Raman responses of the short Be-Se species from Zn$_{1-x}$Be$_x$Se 
single crystals (0.10${\leq}x{\leq}$0.53) and epitaxial layers 
(0$<x{\leq}$0.92) in the context of our `percolation' approach resorting to 
the contour modelling for a support. In the single crystals, the 
strength ratio $R$ between the low- and high-frequency Be-Se modes 
scales as expected for random systems, i.e. as $x/(1-x)$. 
In the epilayers, the $R$ values are larger, corresponding to a 
typical over (sub) representation of $\sim$10\% of the `longer' 
(`shorter') Be-Se bonds from the Be-rich (Zn-rich) region, apparently 
a manifestation of some small degree of CPSO, as earlier discussed 
in relation to GaInP$_2$.$^{13}$ Generally the balance of strength 
between the two modes in the 1-bond$\rightarrow$2-mode TO Raman response 
of each bond species in A$_{1-x}$B$_x$C mixed crystals potentially 
emerges as a straightforward and sensitive quantitative probe 
for CPSO at any $x$ value. In principle, there is no intrinsic 
limit of detection here.

On the theoretical side we perform first-principles calculations 
with a series of fully relaxed Zn$_8$Be$_8$Se$_{16}$ supercells, for 
detailed insight into the dependencies of the vibrational and 
bond length properties on ordering in the representative ZnBeSe$_2$ 
alloy, from the random limit ($\eta$=0) up to full CuPt-type ordering 
($\eta$=1). Two different regimes can be distinguished. 

Regime 1 corresponds to progressive segregation of the Zn and 
Be species within alternate substituting planes, that happens 
so long as the local bonding between the cationic and anionic 
planes keeps locally a (Zn-Se, Be-Se)-mixed character, like 
in a random alloy. Regime 1 leads to `full condensation' of the 
original 1-bond$\rightarrow$2-mode (TO,L) behaviour at $\eta$=0 onto 
the ultimate 1-bond$\rightarrow$1-mode (TO,L) behaviour where the 
Zn-Se and Be-Se modes which eventually survive are those with 
initially closest frequencies, i.e. the harder of the Zn-Se modes 
and the softer of the Be-Se modes. These obviously refer to the 
Zn-Se bonds within the Zn-rich region only, thereby all `short', 
together with the Be-Se bonds within the Be-rich region only, 
thereby all `long', respectively. This way the contrast between 
the Zn-Se and Be-Se bond lengths is minimized, thus optimizing 
the crystal stability. The end of this regime, i.e. the `full 
condensation', is reached at $\eta{\sim}$0.5. In the synthetic scheme 
displayed in Fig.~1, regime 1 is represented by the antagonist curved arrows. 
Basically in the experimental section of this work, we were able to observe 
the onset of regime 1.  

With further local segregation of the Be and Zn species beyond $\eta{\sim}$0.5,
the alloy enters regime 2 corresponding to re-adoption 
of a 1-bond$\rightarrow$2-mode (TO,L) behaviour, the result of the formation 
of ZnSe/BeSe micro-domains here and there. The key point is that 
regime 2 does not seem much rewarding with respect to further 
minimization of the overall contrast between the Zn-Se and Be-Se 
bond lengths in the alloy, which therefore eliminates the driving 
force for a further increase of SO. On the above basis $\eta{\sim}$0.5 
appears as an intrinsic limit to SO in ZnBeSe$_2$, and possibly 
in stoichiometric alloys in general. 

\section*{Acknowledgements}
This work has been supported by the Indo-French Center for 
the Promotion of Advanced Research (IFCPAR project N$^{\circ}$ \mbox{3204-1:} 
\textit{Lattice dynamical and structural study of Be-based 
II-VI semiconductor alloys}), and by an \textit{Action en R\'{e}gion 
de Coop\'{e}ration Universitaire et Scientifique} (ARCUS) programme. The 
authors acknowledge the access to the computing facilities at 
the Centre Informatique National de l'Enseignement Sup\'{e}rieur 
(CINES project N\ensuremath{^\circ}. pli2623, Montpellier, France), and 
the support of the Region Lorraine for awarding one Chaire Internationale 
d'Accueil for A.V. Postnikov at University of Metz. The group 
of Metz would like to thank Dr. K.C. Rustagi for encouragement 
and useful discussions.

\subsection*{References}
\noindent
\begin{tabular}{r@{\hspace*{-0.1mm}}p{17.0cm}} 
$^{~1}$&J.C. Mikkelsen and J.B. Boyce, 
        Phys. Rev. Lett. \textbf{49}, 1412 (1982). \\ 
$^{~2}$&I.F. Chang and S.S. Mitra, 
        Adv. Phys. \textbf{20}, 359 (1971). \\
$^{~3}$&F. Alsina, N. Mestres, J. Pascual, C. Geng, P. Ernst and F. Scholz,
        Phys. Rev. B \textbf{53}, 12994 (1996). \\
$^{~4}$&A. Hassine, J. Sapriel, P. Le Berre, M.A. Di Forte-Poisson, 
        A. Alexandre and M. Quillec, 
	Phys. Rev. B \textbf{54}, 2728 (1996). \\
$^{~5}$&A. Chakrabarti, P. Kratzer and M. Scheffer, 
        Phys. Rev. B \textbf{74}, 245328 (2006). \\
$^{~6}$&A. Mascarenhas, 
        \textit{Spontaneous Ordering in Semiconductor Alloys}, 
	(Kluwer Academics, New York, 2002). \\
$^{~7}$&. S.B. Zhang, S. Froyen and A. Zunger, 
        Appl. Phys. Lett. \textbf{67}, 3141 (1995). \\
$^{~8}$&A. Gomyo, M. Makita, I. Hino and T. Suzuki, 
        Phys. Rev. Lett. \textbf{72}, 673 (1994). \\
$^{~9}$&A. Chakrabarti and K. Kunc, 
        Phys. Rev. B \textbf{72}, 45342 (2005). \\
$^{10}$&Y. Zhang, A. Mascarenhas and L.-W. Wang, 
        Phys. Rev. B \textbf{64}, 125207 (2001). \\
$^{11}$&O. Pag\`{e}s, M. Ajjoun, T. Tite, D. Bormann, E. Tourni\'{e} 
        and K.C. Rustagi, 
	Phys. Rev. B \textbf{70}, 155319 (2004). \\
$^{12}$&A.V. Postnikov, O. Pag\`{e}s and J. Hugel, 
        Phys. Rev. B \textbf{71}, 115206 (2005). \\
$^{13}$&O. Pag\`{e}s, A. Chafi, D. Fristot and A.V. Postnikov, 
        Phys. Rev. B \textbf{73}, 165206 (2006). \\
$^{14}$&H.M. Cheong, F. Alsina, A. Mascarenhas, J.F. Geisz and J.M. Olson, 
        Phys. Rev. B \textbf{56}, 1888 (1997). \\
$^{15}$&F. Alsina, J.D. Webb, A. Mascarenhas, J.F. Geisz, J.M. Olson 
        and A. Duda, 
	Phys. Rev. B \textbf{60}, 1484 (1999). \\
$^{16}$&O. Pag\`{e}s, T. Tite, A. Chafi, D. Bormann, O. Maksimov and 
        M.C. Tamargo, 
	J. Appl. Phys. \textbf{99}, 63507 (2006). \\
$^{17}$&B. Freytag, P. Pavone, U. R\"{o}ssler, K. Wolf, S. Lankes, 
        G. Sch\"{o}tz, A. Naumov, S. Jilka, H. Stanzl and W. Gebhardt, 
        Solid State Commun. \textbf{94}, 103 (1995). \\
$^{18}$&D. Stauffer, \textit{Introduction to Percolation Theory} 
        (Taylor and Francis, London, 1985). \\
$^{19}$&C. V\'{e}ri\'{e}, 
        J. Cryst. Growth \textbf{184/185}, 1061 (1998). \\
$^{20}$&V. Wagner, J.J. Liang, R. Kruse, S. Gundel, M. Keim, A. Waag 
        and J. Geurts, 
	Phys. Stat. Sol. (b) \textbf{215}, 87 (1999). \\
$^{21}$&H.M. Cheong, S. P. Ahrenkiel, M. C. Hanna and A. Mascarenhas, 
        Appl. Phys. Lett. \textbf{73}, 2648 (1998). \\
$^{22}$&M. C. Hanna, H. M. Cheong and A. Mascarenhas, 
        Appl. Phys. Lett. \textbf{76}, 997 (2000). \\
$^{23}$&A. M. Mintairov, P. A. Blagnov, V. G. Melehin, N. N. Faleev, 
        J. L. Merz, Y. Qiu, S. A. Nikishin and A. Temkin, 
        Phys. Rev. B \textbf{56}, 15836 (1997). \\
$^{24}$&N. Mestres, F. Alsina, J. Pascual, J.M. Bluet, J. Camassel, 
        C. Geng and F. Scholz, 
	Phys. Rev. B \textbf{54}, 17754 (1996). \\
$^{25}$&C. Chauvet, Ph.D thesis, 
        Centre de Recherche sur l'H\'{e}t\'{e}ro\'{e}pitaxie 
        et ses Applications (CNRS), University of Nice Sophia-Antipolis, 
	2001. \\
$^{26}$&O. Pag\`{e}s, M. Ajjoun, D. Bormann, C. Chauvet, E. Tourni\'{e}, 
        J.P. Faurie and O. Gorochov, 
	J. Appl. Phys. \textbf{91}, 9187 (2002). \\
$^{27}$&O. Pag\`{e}s, M. Ajjoun, J.P. Laurenti, D. Bormann, C. Chauvet, 
        E. Tourni\'{e}, J.P. Faurie and O. Gorochov, 
	Opt. Mat. \textbf{17}, 323 (2001). \\
$^{28}$&R. Loudon, 
        Adv. Phys. \textbf{13}, 423 (1964). \\
$^{29}$&P. Ordejon, E. Artacho and J. M. Soler, 
        Phys. Rev. B \textbf{53}, R10441 (1996). \\
$^{30}$&J. M. Soler, E. Artacho, J. D. Gale, A. Garcia, J. Junquera, 
        P. Ordejon, and D. Sanchez-Portal, 
	J. Phys.: Condens. Matt. \textbf{14}, 2745 (2002). \\
$^{31}$&A. Bukaluk, A.A. Wronkowska, A. Wronkowski, H. Arwin, F. Firszt, 
        S. {\L}\k{e}gowski, H. M\k{e}czy\'{n}ska, J. Szatkowski, 
	Appl.  Surf. Sci. \textbf{175-176}, 531 (2001). \\
$^{32}$&C. Chauvet, E. Tourni\'{e} and J.P. Faurie, 
        Phys. Rev. B \textbf{61}, 5332 (2000). \\
$^{33}$&E. Tourni\'{e}, C. Morhain, G. Neu, M. La\"{u}gt, C. Ongaretto, 
        J.-P. Faurie, R. Triboulet and J. O. Ndap, 
	J. Appl. Phys. \textbf{80}, 2983 (1996). \\
$^{34}$&M.-H. Tsai, F. C. Peiris, S. Lee and J. K. Furdyna, 
        Phys. Rev. B \textbf{65}, 235202 (2002). \\
$^{35}$&T. Tite, O. Pag\`{e}s, E. Tourni\'{e}, 
        Appl. Phys. Lett. \textbf{85}, 5872 (2004). \\
$^{36}$&O. Pag\`{e}s, T. Tite, K. Kim, P.A. Graf, O. Maksimov 
        and M.C. Tamargo, 
	J. Phys.: Condens. Matter \textbf{18}, 577 (2006). \\
$^{37}$&A. Silverman, A. Zunger, R. Kalish and J.Adler, 
        Phys. Rev. B \textbf{51}, 10795 (1995). \\
\end{tabular}

\end{document}